\documentclass[preprint, 
amsmath, amssymb, superscriptaddress, 
nofootinbib, 
showpacs,
preprintnumbers, keywords]{revtex4}

\usepackage{bm}
\usepackage{booktabs}
\usepackage{mathrsfs}
\usepackage{braket}
\usepackage{slashed}

\usepackage[dvipdfmx]{graphicx}

\usepackage{graphicx,color}
\usepackage[utf8]{inputenc}
\usepackage{color}
\usepackage{ulem}

\newcommand{\Slash}[1]{{\ooalign{\hfil#1\hfil\crcr\raise.167ex\hbox{/}}}}
\usepackage{fancybox} 
\usepackage{multirow}

\begin{document}

\title{
Determination of coupling patterns by parallel searches for $\mu^-\to e^+$ and $\mu^-\to e^-$ in muonic atoms
}

\author{Joe Sato}
\email[E-mail: ]{sato-joe-mc@ynu.ac.jp}
\affiliation{Department of Physics, Graduate School of Engineering Science,
Yokohama National University, Yokohama, 240-8501, Japan}

\author{Kohei Sugawara}
\email[E-mail: ]{sugawara@krishna.th.phy.saitama-u.ac.jp}
\affiliation{Physics Department, Saitama University, 
Saitama 338-8570, Japan}

\author{Yuichi Uesaka}
\email[E-mail: ]{uesaka@ip.kyusan-u.ac.jp}
\affiliation{Faculty of Science and Engineering, 
Kyushu Sangyo University, 2-3-1 Matsukadai, Higashi-ku, 
Fukuoka 813-8503, Japan}

\author{Masato Yamanaka}
\email[E-mail: ]{yamanaka@osaka-cu.ac.jp}
\affiliation{Department of Mathematics and Physics, 
Osaka City University, Osaka 558-8585, Japan}
\affiliation{Nambu Yoichiro Institute of Theoretical and 
Experimental Physics (NITEP), Osaka City University, 
Osaka 558-8585, Japan}

\preprint{\bf STUPP-21-254, OCU-PHYS 559, NITEP 131}

\date{\today}

\begin{abstract}
We investigate a possibility that the $\mu^-\to e^+$ conversion is discovered 
prior to the $\mu^-\to e^-$ conversion, and its implications to the new physics 
search. We focus on the specific model including the mixing of the $SU(2)_L$ doublet- 
and singlet-type scalar leptoquarks, which induces not only the lepton flavor violation 
but also the lepton number violation. Such a structure is motivated by R-parity 
violating (RPV) supersymmetric models, where a sbottom mediates the conversion 
processes. We formulate the $\mu^-\to e^+$ rate in analogy with the muon capture 
in a muonic atom, and numerically evaluate it using several target nuclei. The lepton 
flavor universality test of pion decay directly limits the $\mu^-\to e^+$ rate, 
and the maximally allowed $\mu^-\to e^+$ branching ratio is $\sim 10^{-18}$ 
under the various bounds on RPV parameters. 
We show that either $\mu^-\to e^-$ or $\mu^-\to e^+$ signals can be 
discovered in near future experiments. This indicates that parallel searches for these 
conversions will give us significant information on the pattern of coupling constants.
\end{abstract}

\maketitle

\section{Introduction}
\label{sec:Intro}

The standard model (SM), where all neutrinos are left-handed and massless, 
has the accidental global $U(1)$ symmetries which ensure to conserve 
the lepton flavor numbers, $L_e$, $L_\mu$, and $L_\tau$.  
Nonetheless, the lepton flavor violation (LFV) was established by the 
discovery of neutrino oscillation, which implies that the three global 
symmetries are broken and the SM should be extended to include LFV sources.

In lots of extended models, LFV sources cause not only 
the flavor violation among charged leptons (called CLFV) but also 
the lepton number violation (LNV). 
One may presume that the LNV processes are minor 
compared with CLFV, because, aside from the flavor number, the 
particle number must be violated. 
However, we know situations where it does not hold. 
A well-known example is the Majorana mass of the neutrinos; the branching ratio of an LNV process 
$\mu^- \to e^+$ in nuclei could be much larger than that of LFV 
process $\mu \rightarrow e\gamma$ due to the GIM suppression in 
the flavor changing neutral current\,\cite{Marciano:1977wx, 
Bilenky:1977du,Lee:1977tib}. 
Therefore, both the LFV and LNV processes should be investigated.

The muonic atom is a good probe to both the LFV and LNV; an LFV process 
$\mu^-\to e^-$ conversion, $\mu^-(Z,A)\to e^-(Z,A)$, and an LNV 
process $\mu^-\to e^+$ conversion, $\mu^-(Z,A)\to e^+(Z-2,A)$.  
See Ref.~\cite{Lee:2021hnx} for the recent review of the $\mu^-\to e^+$ conversion.
The experimental signals of these modes is single monoenergetic electron 
(positron), which is highly clean signal with little SM background. 
In near future experiments, the searches for these modes are planned by 
using a number of muonic atoms (COMET \cite{Abramishvili_2020}, 
Mu2e \cite{Bartoszek:2014mya}, and PRISM/PRIME \cite{Barlow:2011zza}).

In this article, we investigate a possibility that the $\mu^-\to e^+$ conversion 
could be discovered prior to the $\mu^-\to e^-$ conversion. 
An interesting example to address the possibility is leptoquarks with the mixing of 
$SU(2)$ doublet and singlet.
The condition is satisfied by sbottoms in R-parity violating (RPV) supersymmetric 
(SUSY) model~\cite{Babu:1995vh}. When the sbottom $\tilde{b}$ has the RPV 
interaction $\tilde{b}\ell q$ and the mixing of $SU(2)$ doublet $\tilde{b}_L$ 
and singlet $\tilde{b}_R$, the lepton number is not conserved and the 
$\mu^-\to e^+$ conversion can be induced at tree level. 
We formulate the $\mu^-\to e^+$ conversion rate for the sbottom mediation, 
and numerically evaluate it under the experimental bounds on RPV parameters. 
We see that importance to search for and analyze the non-standard reactions 
of muonic atoms without prejudice that the LFV reactions are always leading compared with the LNV ones.

The contents of this article are as follows:
In Sec.~\ref{Sec:model}, we introduce leptoquarks inspired by sbottom in RPV SUSY 
and discuss current constraints on the coupling constants.
We show the formula for the rate of the $\mu^- \to e^-$ and $\mu^- \to e^+$ 
conversions in a muonic atom in Sec.~\ref{Sec:muonic_atom}.
The results are shown in Sec.~\ref{Sec:Result}, and finally, the article is summarized 
in Sec.~\ref{sec:Summary}.

\section{Benchmark model}
\label{Sec:model}

We introduce a benchmark SUSY model wherein the reaction rates of 
$\mu^-\to e^+$ conversion and $\mu^-\to e^-$ conversion are 
comparable to each other.

The gauge invariant superpotential contains the RPV terms 
\cite{Weinberg:1981wj, Sakai:1981pk, Hall:1983id}, and one of them could 
be a source of LFV, $\mathcal{W}_\text{RPV} = \lambda'_{ijk} L_i Q_j D_k^c$. 
Here $D_i$ is a $SU(2)_L$ singlet superfield, and $L_i$ and $Q_i$ are $SU(2)_L$ 
doublet superfields. Indices $i$, $j$, and $k$ represent the generations. 
The interaction terms related with LFV and LNV processes are 
\begin{equation}
\begin{split}
   &\mathcal{L}_{\lambda'}
   = \lambda'_{ijk}  \left[ 
   \widetilde d_{jL} \overline{d}_{kR} \nu_{iL} 
   - \widetilde d_{kR}^* \overline{(e_{iL})^c} u_{jL}
   \right] + \text{H.c.},
\label{Eq:RPV_L1}   
\end{split}      
\end{equation}
where $\tilde{d}_j$ is the SUSY partner of down-type quark $d_j$. 
We assume the simple situation that only the lighter sbottom contributes to 
low-energy observables, which is motivated by that, in many SUSY 
scenarios, it is lighter than the first and second generation 
squarks\,\cite{Martin:1997ns}. 
Thus, $j$ ($k$) in $\widetilde d_{jL}$ ($\widetilde d_{kR}^*$) must be 3.
The left- and right-handed sbottom ($\tilde{b}_L$ and $\tilde{b}_R$) 
are mixed each other after the $SU(2)_L$ symmetry breaking, and it 
could be large as $m_{LR}^2 \propto m_{b} (A_b - \mu \tan\beta)$. 
Here $A_b$ is so-called the trilinear scalar coupling, $\mu$ is the 
higgsino mass parameter, and $\tan\beta$ is the ratio of Higgs field vevs. 
The mixing is parametrized through the diagonalization of sbottom 
mass as 
\begin{align}
\label{Eq:Sbottommatrix}
-\mathcal{L}_{\tilde{b}\text{-}\mathrm{mass}}=
        \left(\tilde{b}_L^* \ \ \tilde{b}_R^* \right)
       \left(
		\begin{array}{cc}
		  m_L^2 & m_{LR}^2  \\
		  m_{RL}^2   & m_R^2 \\
		\end{array}
	\right)
       \left(
		\begin{array}{c}
		  \tilde{b}_L  \\
		  \tilde{b}_R  \\
		\end{array}
	\right)
=
        \left( \tilde{b}_1^* \ \ \tilde{b}_2^*	\right)
        \left(
		\begin{array}{cc}
		  m_1^2 & 0  \\
		  0   & m_2^2 \\
		\end{array}
	\right)
       \left( 
		\begin{array}{c}
		  \tilde{b}_1  \\
		  \tilde{b}_2  \\
		\end{array}
	\right),
\end{align}
where we set $m_1\le m_2$ and take the mixing angle $\theta_{\tilde{b}}$ as
\begin{align}\label{Eq:Sbottommasseigenstate}
       \left(
		\begin{array}{c}
		  \tilde{b}_1  \\
		  \tilde{b}_2  \\
		\end{array}
	\right)
=
       \left(
		\begin{array}{cc}
		  {\rm cos}\theta_{\tilde{b}} & -{\rm sin}\theta_{\tilde{b}}  \\
		  {\rm sin}\theta_{\tilde{b}} & {\rm cos}\theta_{\tilde{b}} \\
		\end{array}
	\right)
       \left(
		\begin{array}{c}
		  \tilde{b}_L  \\
		  \tilde{b}_R  \\
		\end{array}
	\right).
\end{align}
Thus the RPV interaction Lagrangian in terms of mass eigenstates is 
\begin{align}
   &\mathcal{L}_{\lambda'}
   \supset \tilde{\lambda}'_{i31}
    \widetilde b_1 \overline{d}_{R} \nu_{iL} 
   + \tilde{\lambda}'_{i13}\widetilde b_1^* \overline{(e_{iL})^c} u_{L}
    + \text{h.c.}.
\label{Eq:RPVLmassbase}   
\end{align}
where we define $\tilde{\lambda}'_{i31}=\lambda'_{i31} 
\cos\theta_{\tilde{b}}$ and $\tilde{\lambda}'_{i13}=\lambda'_{i13}
\sin\theta_{\tilde{b}}$.

\begin{figure}[ht]
\centering
\includegraphics[keepaspectratio, width=0.4\hsize]{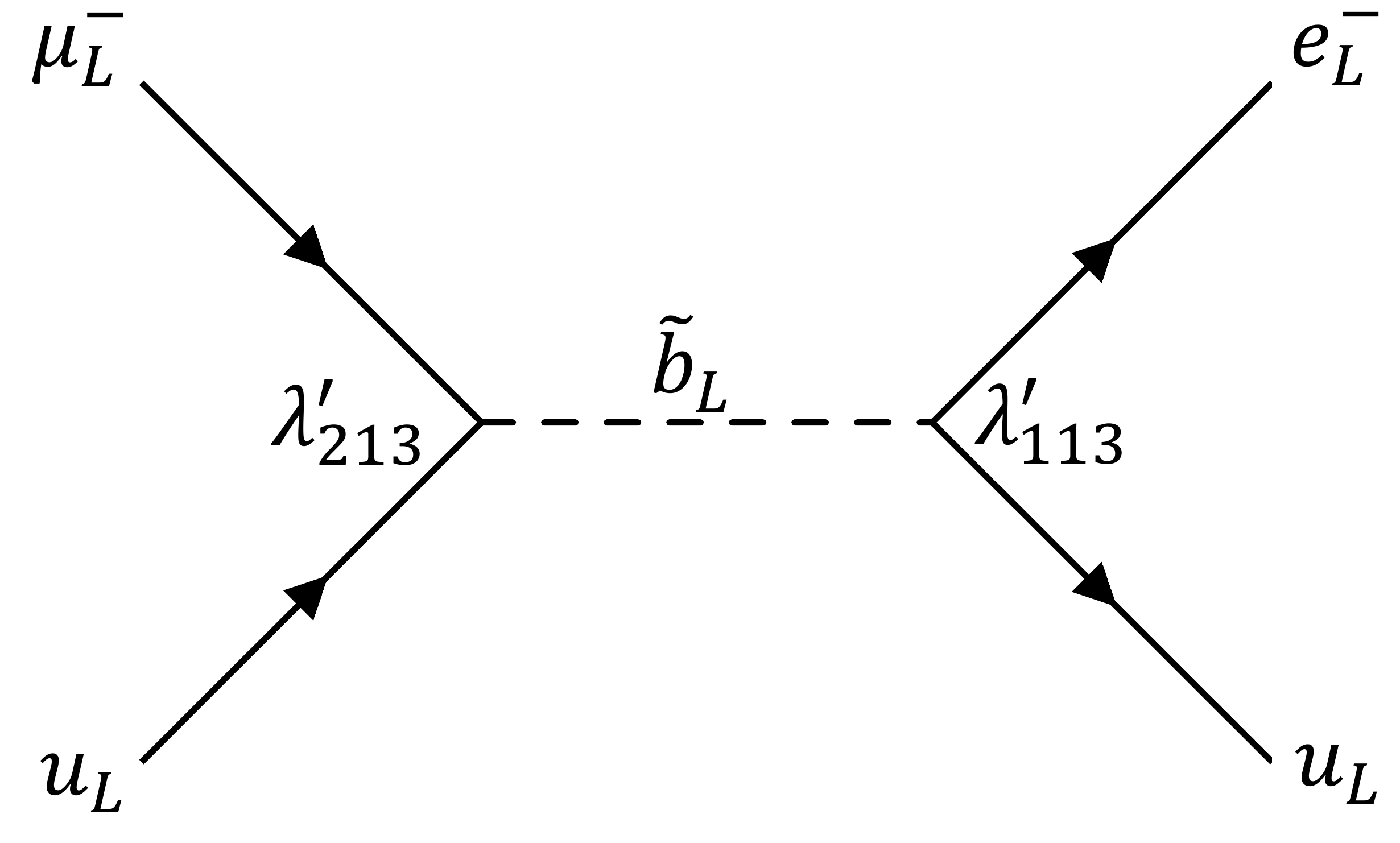}
\caption{$\mu^-\to e^-$ conversion via the RPV operator, Eq.~\eqref{Eq:RPV_L1}.}
\label{fig:diagram_mu2e-}
\end{figure}

\begin{figure}[ht]
\begin{minipage}[t]{0.49\hsize}
\centering
\includegraphics[keepaspectratio, width=0.8\hsize]{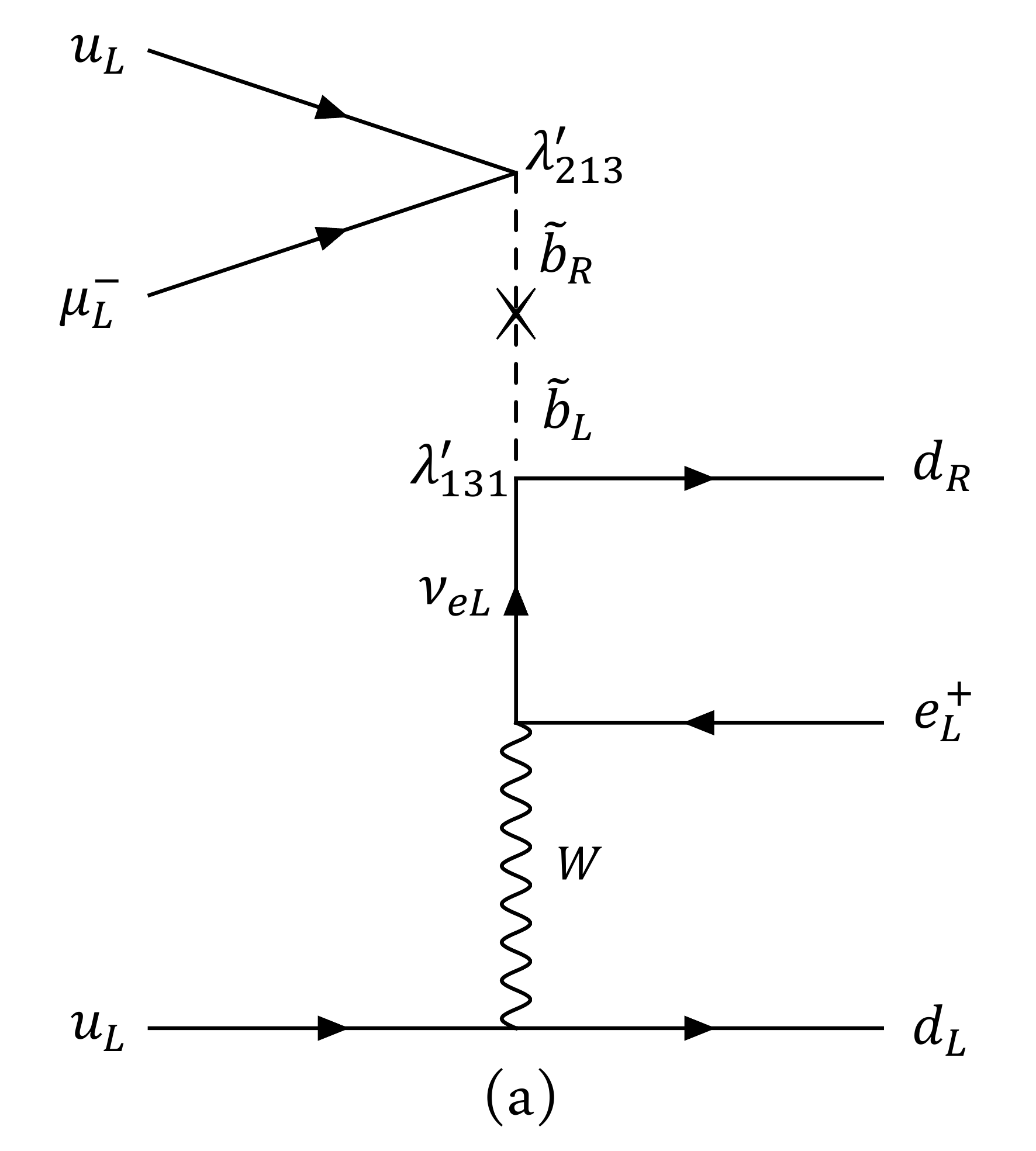}
\end{minipage} 
\begin{minipage}[t]{0.49\hsize}
\centering
\includegraphics[keepaspectratio, width=0.8\hsize]{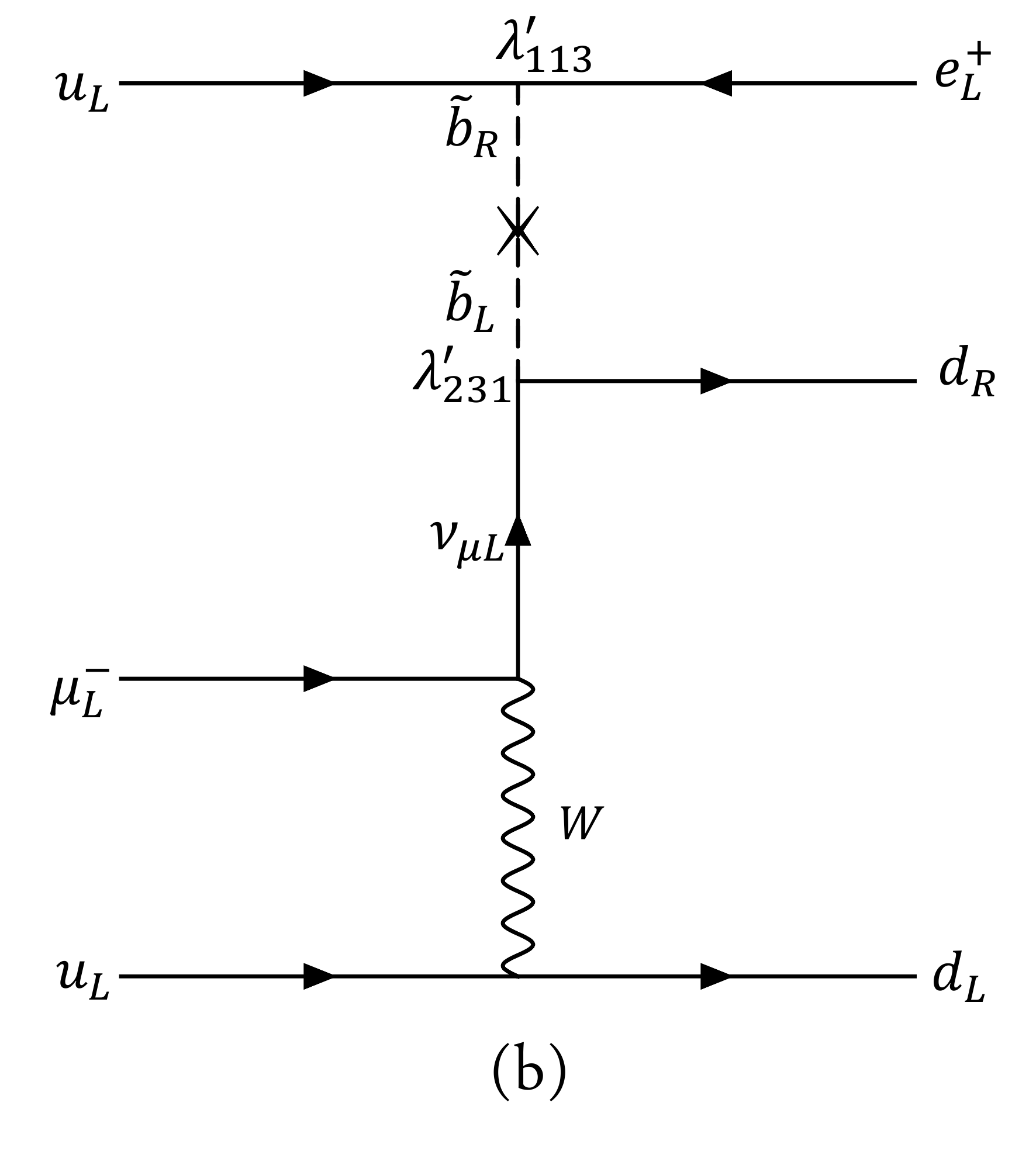}
\end{minipage}
\raggedleft
\caption{$\mu^-\to e^+$ conversion by (a) the combination of 
$\{ \lambda'_{213}, \lambda'_{131} \}$ and (b) the combination 
of $\{ \lambda'_{113}, \lambda'_{231} \}$.}
\label{fig:RPV}
\end{figure}

The lepton flavors are no longer defined as conserved quantities with the interactions in Eq.~\eqref{Eq:RPV_L1}.
Then, the $\mu^- \to e^-$ conversion in nuclei is induced by the exchange of $\tilde{b}_L$ as shown in Fig.~\ref{fig:diagram_mu2e-}.

When the $\tilde{b}_L$-$\tilde{b}_R$ mixing exists in addition to the RPV interactions, the lepton number conservation is violated:
if the mixing is absent, the lepton number $-1$ ($+1$) can be assigned to $\tilde{b}_L$ ($\tilde{b}_R$).
The $\mu^- \to e^+$ conversion in nuclei 
arises via the LFV vertex and the $\tilde{b}_L$-$\tilde{b}_R$ mixing 
(Fig.\,\ref{fig:RPV}).
It is important to emphasize that, when either $\lambda'_{213}$ or 
$\lambda'_{113}$ is zero, the $\mu^-\to e^-$ conversion rate goes 
to zero, but the $\mu^-\to e^+$ conversion could be observable.

The experimental bounds on the RPV parameters are set by independent 
measurements. We summarize the bounds in the rest of this section.

\subsection{Atomic parity violation and parity violating electron scattering}
\label{sec:APV_PVES}

The measurements of atomic parity violation (APV) and parity 
violating electron scattering (PVES) test the parity violating interaction, 
and set the bound on $\lambda'_{131}$~\cite{Barger:1989rk}.
The parity violating interaction is parametrized as 
$-(G_F/\sqrt{2}) C_{1i} \bar{e} \gamma_\mu \gamma_5 e 
\bar{q}_i \gamma^\mu q_i$, where $G_F=1.166 \times 
10^{-5}$\,GeV$^{-2}$ is the Fermi coupling constant.
The sbottom interferes with the photon and $Z$ boson in APV and 
PVES, and the effective coupling is\footnote{We neglect 
the QED corrections to the $C_{1d}$ because it is small, 
$\left| C_{1d}^{\rm{w}} - C_{1d}^{\rm{w/o}} \right| 
/C_{1d}^{\rm{w/o}} \simeq 
\mathcal{O}(1)$\,\cite{Amaldi:1987fu}, and the 
resultant effect on the $\lambda'_{131}$ bound is negligible.}
$C_{1d} = \dfrac{1}{2} - \dfrac{2}{3} \sin^2 \theta_w 
+ \dfrac{m_W^2}{g^2} 
\dfrac{\bigl| \tilde{\lambda}'_{131} \bigr|}{m_{\tilde{t}_L}^2}$. 
$C_{1d}$ is obtained by including the APV results in the global 
fit incorporating the Qweak collaboration result and PVES database, 
$C_{1d} = 0.3389 \pm 0.0025 \ (1 \sigma)$\,\cite{Androic:2018kni}.
With $\sin^2 \theta_w = 0.2382$ at the experimental scale, the bound is 
\begin{equation}
\begin{split}
	\bigl| \tilde{\lambda}'_{131} \bigr|
	\leq 6.9 \times 10^{-1} 
	\left( \frac{m_{\widetilde{t}_L}}{1\,\rm{TeV}} \right),
	\label{Eq:bound-131}
\end{split}
\end{equation}
which depends on the assumption of the stop mass $m_{\tilde{t}_L}$.
If the stop is sufficiently heavy, substantially there is no constraint on the coupling.
In the analysis of this article, we will set $m_{\tilde{t}_L}=1$ TeV to 
have a bound, $\bigl| \tilde{\lambda}'_{131} \bigr|<0.69$.

\subsection{Neutrino-nucleon scattering: $\nu_\mu d_R \to \nu_\mu d_R$}
\label{sec:DIS}

The sbottom exchange subprocess via $\lambda'_{231}$ interferes 
with the SM neutrino deep inelastic scattering (DIS) $\nu_\mu d_R \to 
\nu_\mu d_R$~\cite{Barger:1989rk}. 
Taking into account the interference, the coupling for the neutral 
current connecting $\nu_\mu$ and $d_R$ is 
$g_R^d = \dfrac{1}{3} \sin^2\theta_W + \dfrac{m_W^2}{g^2} 
\dfrac{\bigl| \tilde{\lambda}'_{231} \bigr|^2}{m_{1}^2}$. 
The precision measurement of the neutrino DIS provides 
$g_R^d = -0.027^{+0.077}_{-0.048}$~\cite{ParticleDataGroup:2020ssz}, 
which excludes nonzero $\lambda'_{231}$ at the $1 \sigma$ level. 
The bound at the $2 \sigma$ level is 
\begin{equation}
\begin{split}
	\bigl| \tilde{\lambda}'_{231} \bigr|
	\leq 3.6 \times 10^{-1} 
	\left( \frac{m_1}{200\,\rm{GeV}} \right). 
	\label{Eq:bound-231}
\end{split}
\end{equation}

\subsection{Direct sbottom search}
\label{sec:sbottom_decay}

The direct search sets the limits on sbottom mass and RPV couplings. 
The decay width of RPV channel $\tilde{b}_1 \to e_{lL} u_L$ is 
\begin{align}
	\Gamma \bigl( \tilde{b}_1 \to e_{lL} u_L \bigr) 
	=& \frac{\bigl| \tilde{\lambda}_{l13}^\prime \bigr|^2}{16\pi m_1} 
	\lambda \left(1,\frac{m_{e_{l}}^2}{m_1^2},\frac{m_u^2}{m_1^2}\right) 
	\left( m_1^2-m_{e_{l}}^2-m_u^2 \right).
\label{Eq:neutralinobound2}
\end{align}
Here $\lambda(x,y,z)=\sqrt{x^2+y^2+z^2-2xy-2yz-2zx}$. The decay 
width of R-parity conserving channel $\tilde{b}_1 \to \tilde{\chi}^0 b$ is 
\begin{align}
	&\Gamma \bigl( \tilde{b}_1 \to \tilde{\chi}^0 b \bigr) 
	= \frac{g_1^2}{16\pi m_1} 
	\lambda \biggl(1,\frac{m_{\tilde{\chi}^0}^2}{m_1^2}, 
	\frac{m_b^2}{m_1^2} \biggr) \nonumber 
	\\& \hspace{5mm} \times 
	\Bigl[  
	\left( Y_L^2\cos^2\theta_{\tilde{b}}+Y_R^2\sin^2\theta_{\tilde{b}} \right)
	\left( m_1^2-m_{\tilde{\chi}^0}^2-m_b^2 \right) 
	-8Y_L Y_R \sin \theta_{\tilde{b}} \cos \theta_{\tilde{b}} m_b m_{\tilde{\chi}^0} 
	\Bigr],
\label{Eq:neutralinobound}
\end{align}
where $Y_L$ and $Y_R$ are the hypercharge for left- and right-handed 
bottom, $m_{\tilde{\chi}^0}$ is the neutralino mass, and $m_b$ is the 
bottom mass. 
Setting the mass scales by maximally small ones $m_1=200$\,GeV and 
$m_{\tilde{\chi}_0}=160$\,GeV\,\cite{ATLAS:2017avc}, the direct 
search limit $\Gamma \bigl( \tilde{b}_1 \to e_{lL} u_L \bigr)/
\Gamma \bigl( \tilde{b}_1 \to \tilde{\chi}^0 b \bigr) < \mathcal{O} 
\left(10^{-2}\right)$ ($l=e,\mu$)\,\cite{ATLAS:2019ebv, ATLAS:2021hza} 
is transferred to the bound on RPV coupling as 
\begin{align}
	\bigl| \tilde{\lambda}'_{i13} \bigr| \lesssim 5 \times 10^{-3}. 
\label{Eq:direct}
\end{align}

\subsection{Lepton flavor universality of pion decays}
\label{sec:lepton_universality}

\begin{figure}[ht]
  \begin{minipage}[b]{0.24\linewidth}
    \centering
    \includegraphics[width=4cm]{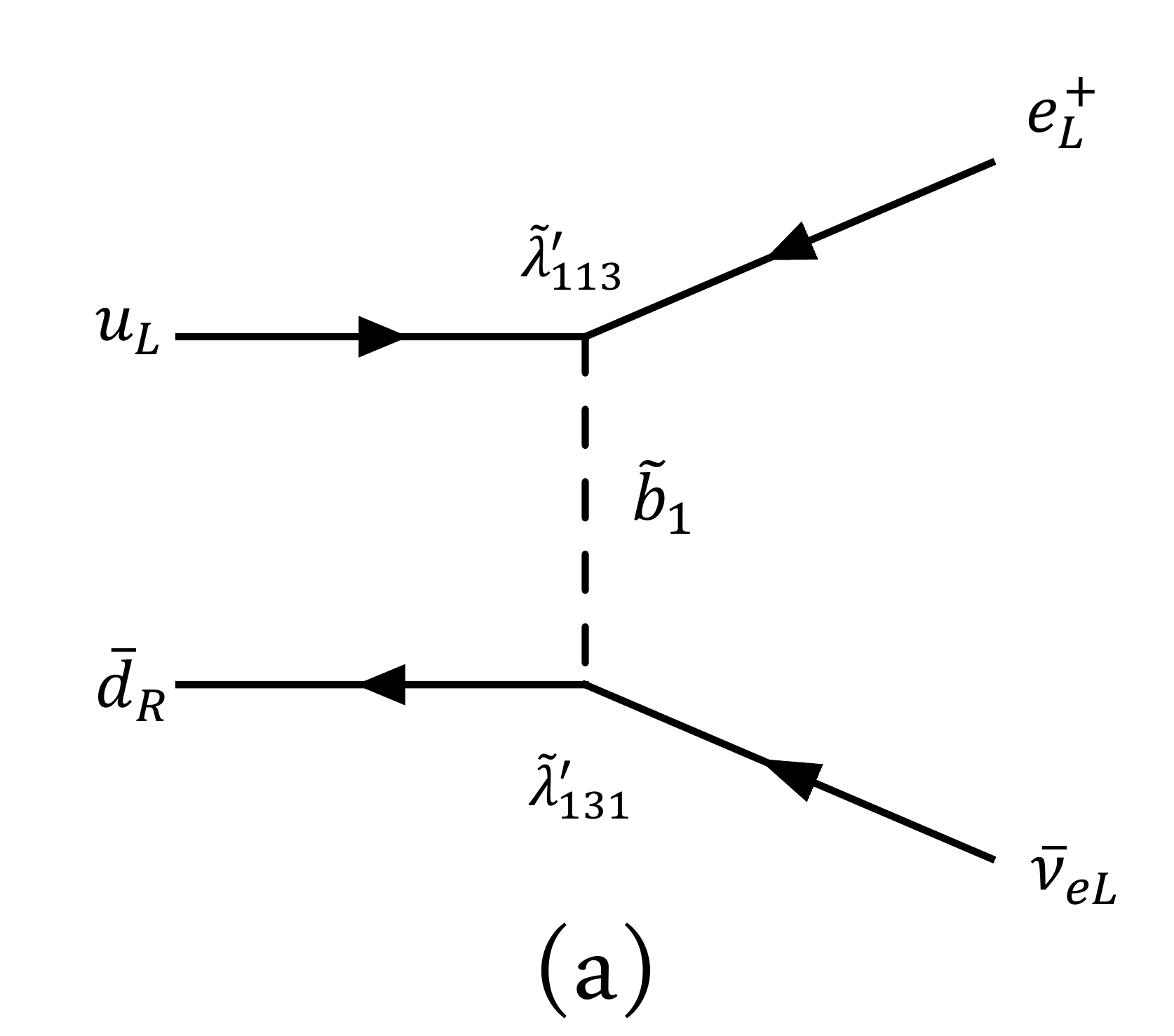}
  \end{minipage}
  \begin{minipage}[b]{0.24\linewidth}
    \centering
    \includegraphics[width=4cm]{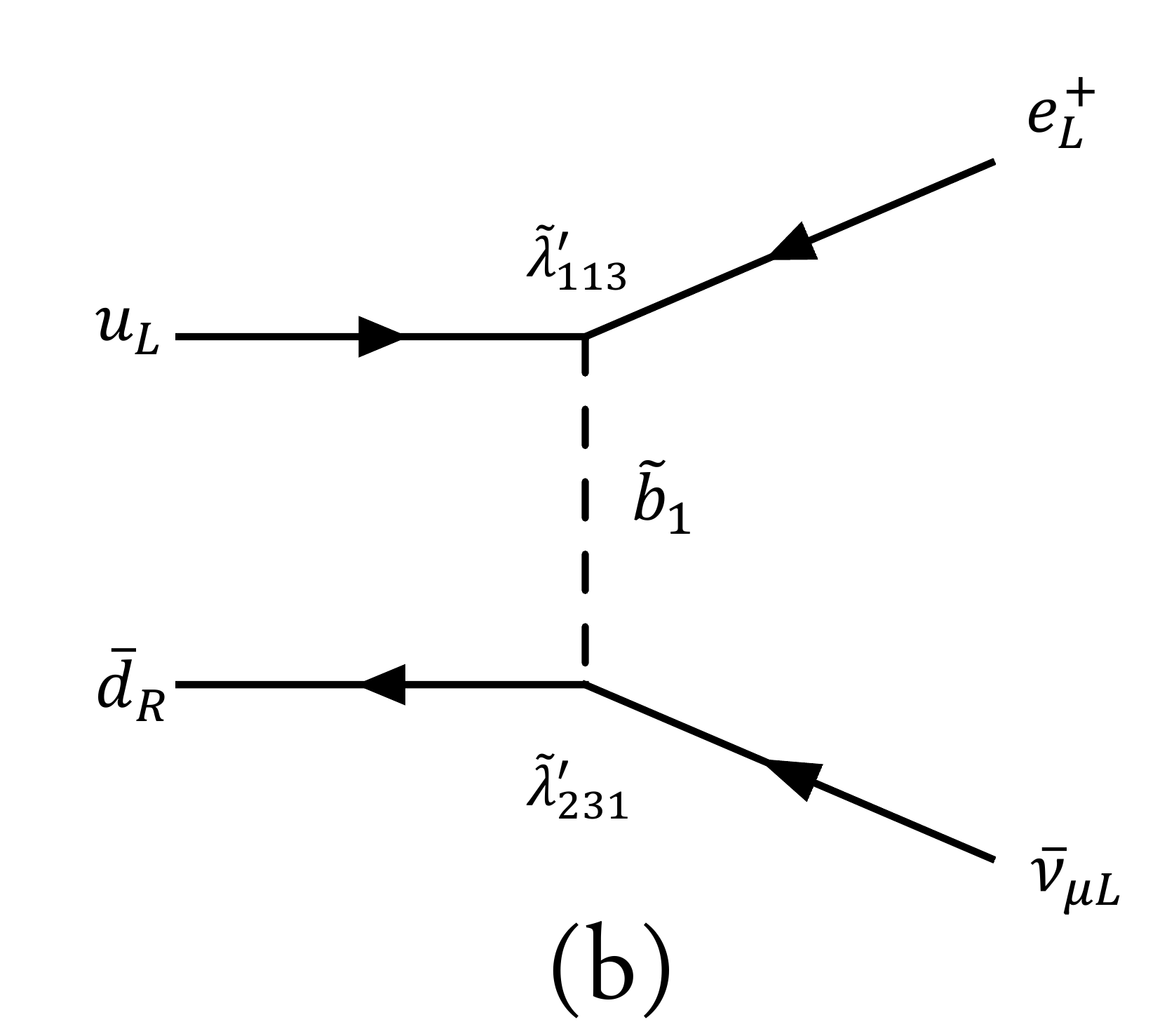}
  \end{minipage}
  \begin{minipage}[b]{0.24\linewidth}
    \centering
    \includegraphics[width=4cm]{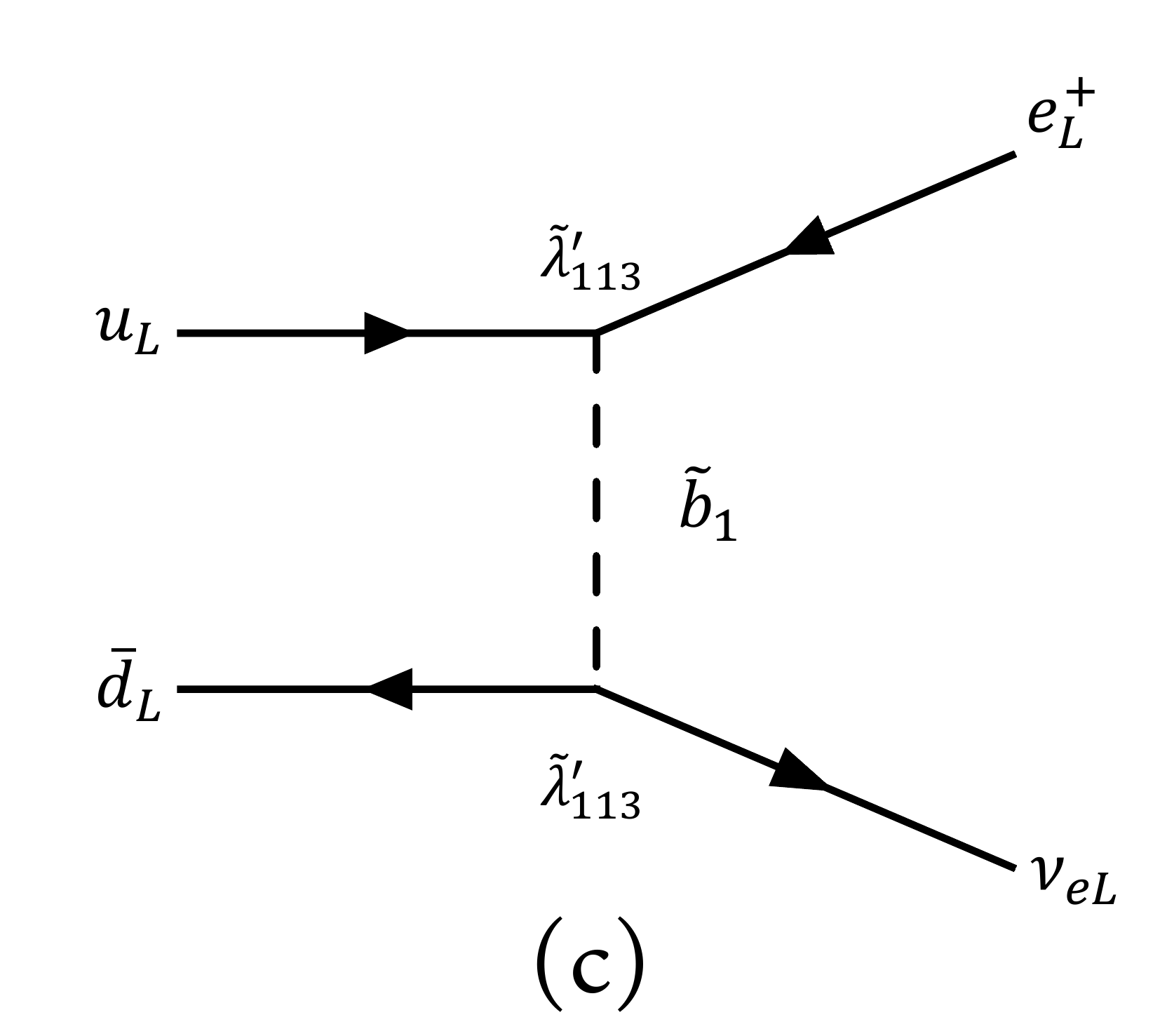}
  \end{minipage}
  \begin{minipage}[b]{0.24\linewidth}
    \centering
    \includegraphics[width=4cm]{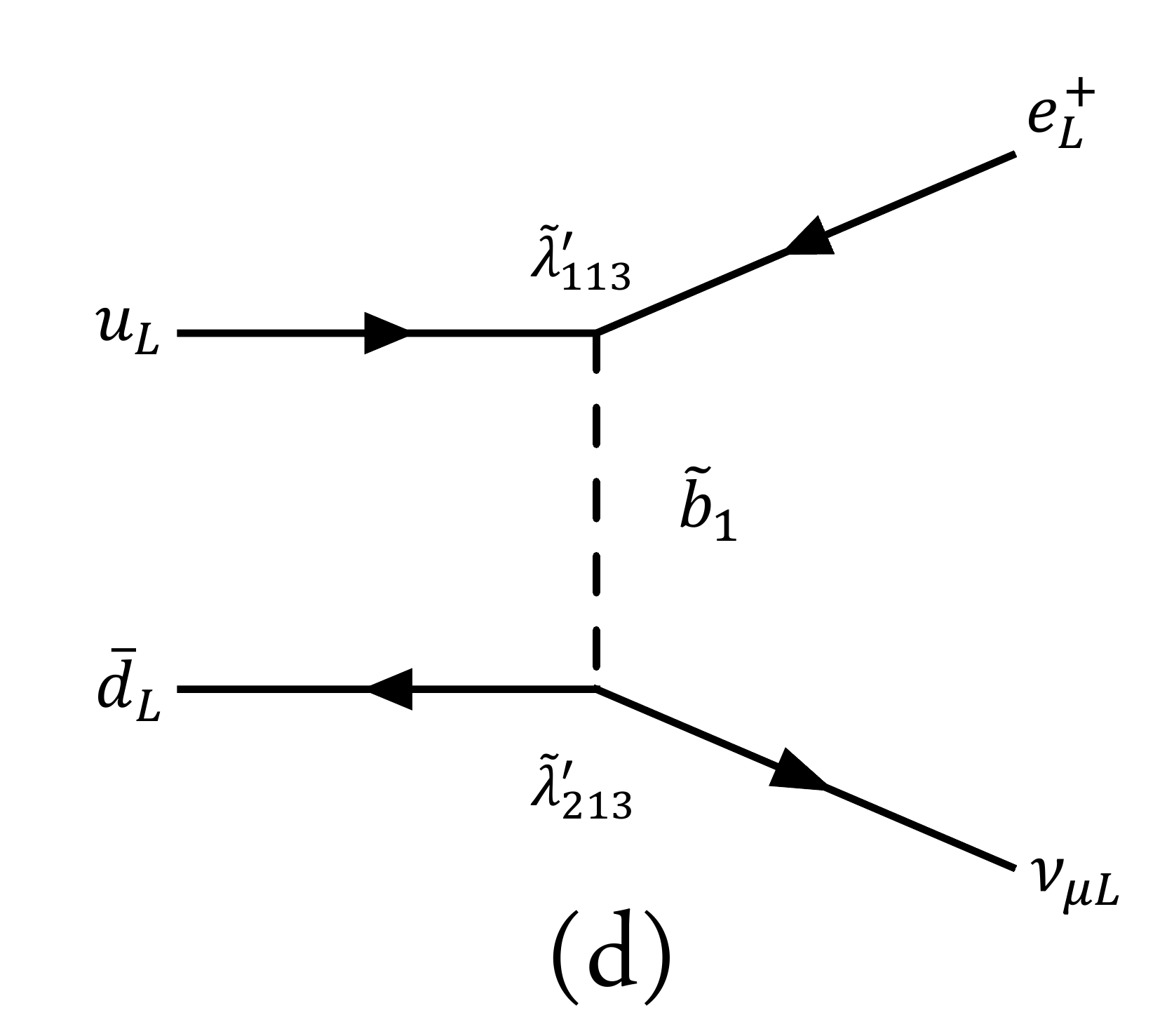}
  \end{minipage}
  \begin{minipage}[b]{0.24\linewidth}
    \centering
    \includegraphics[width=4cm]{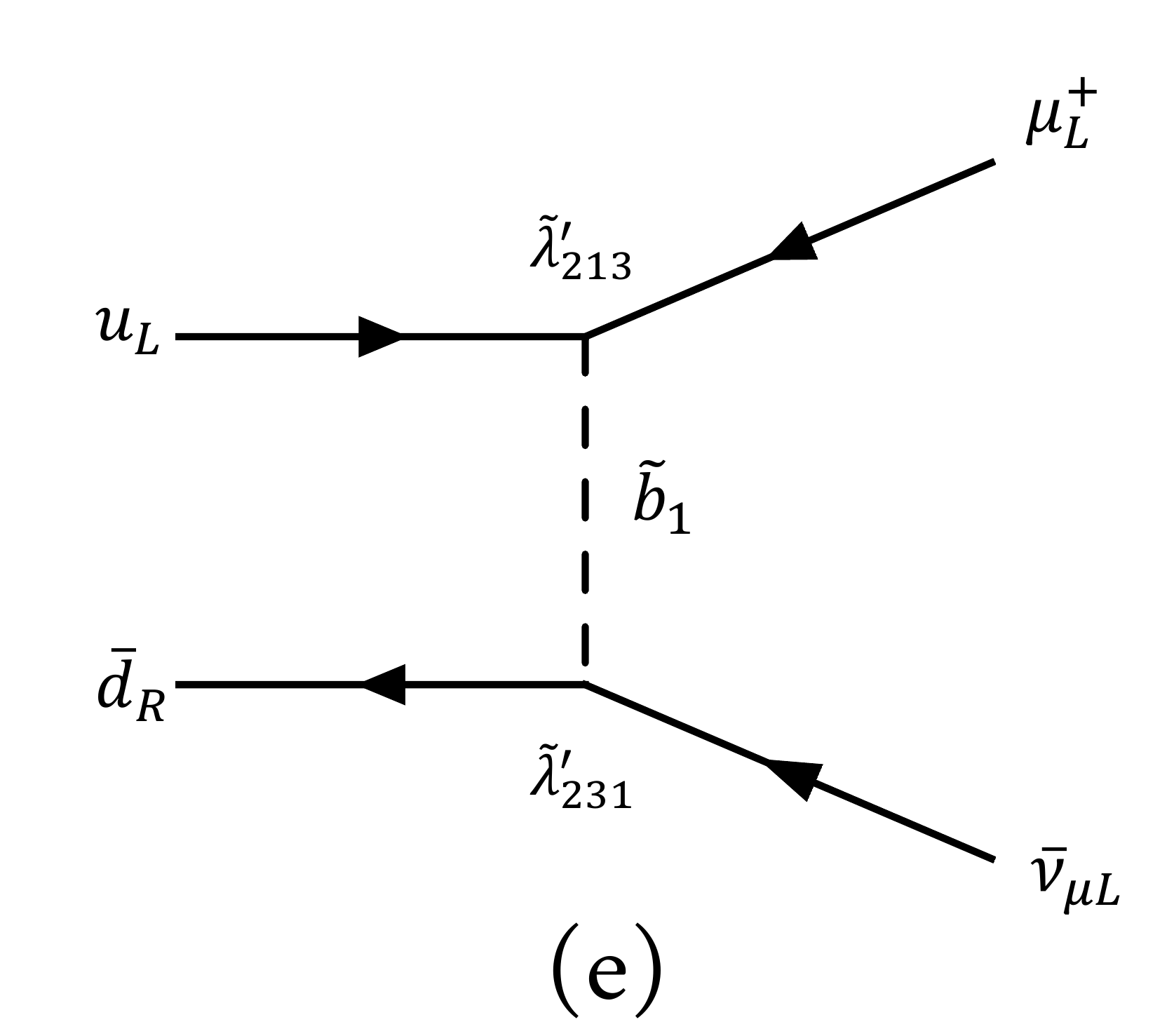}
  \end{minipage}
  \begin{minipage}[b]{0.24\linewidth}
    \centering
    \includegraphics[width=4cm]{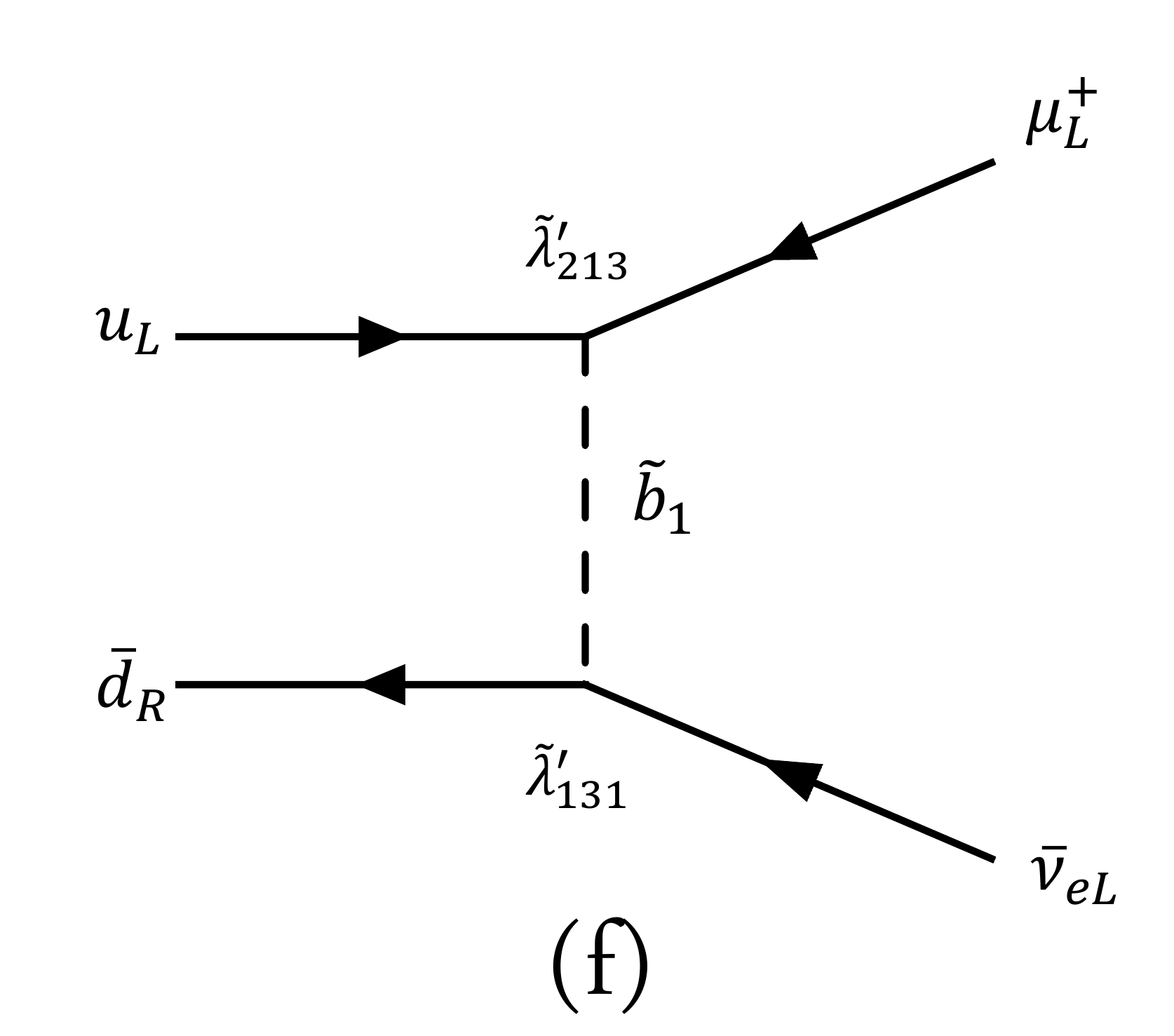}
  \end{minipage}
  \begin{minipage}[b]{0.24\linewidth}
    \centering
    \includegraphics[width=4cm]{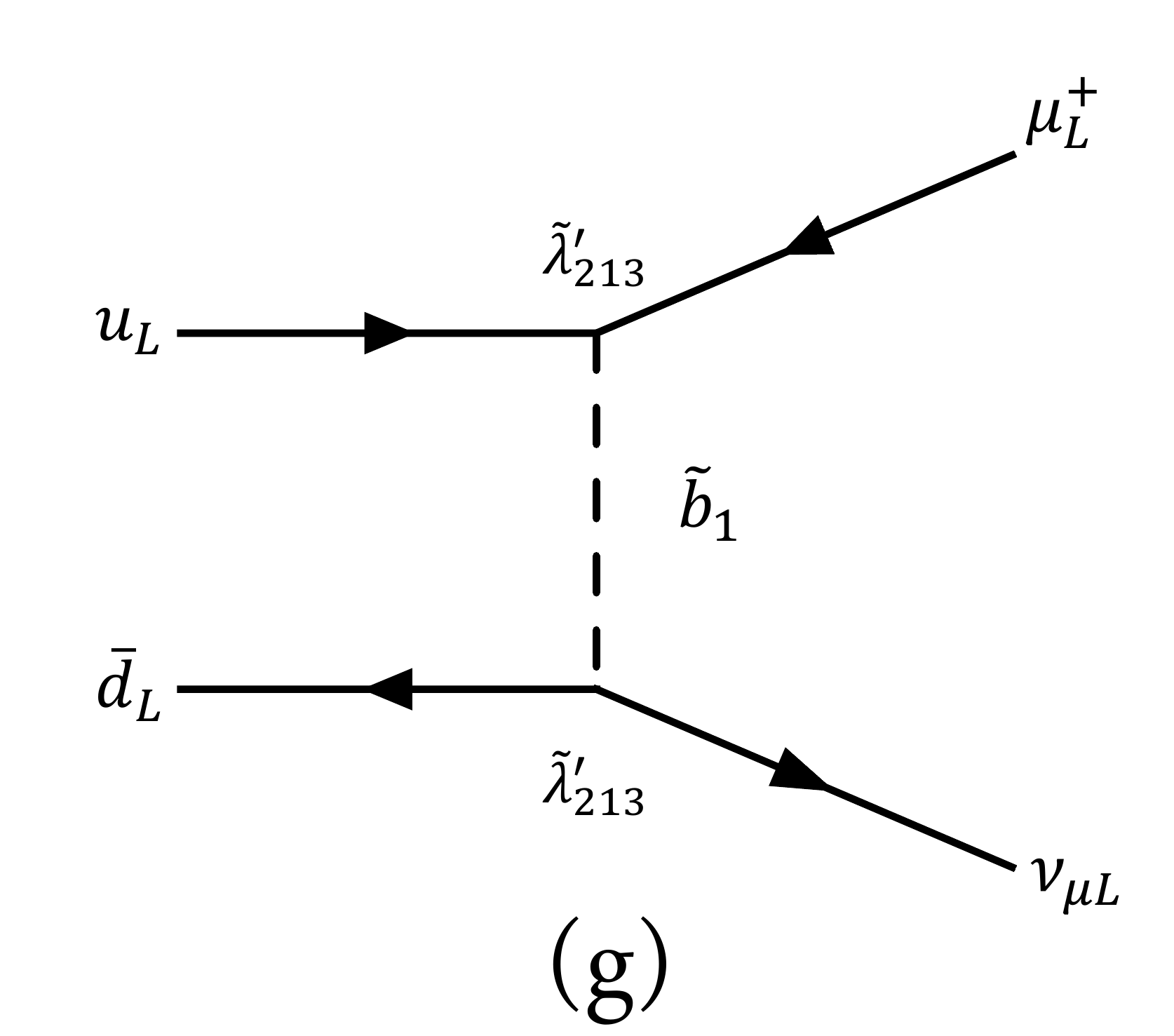}
  \end{minipage}
  \begin{minipage}[b]{0.24\linewidth}
    \centering
    \includegraphics[width=4cm]{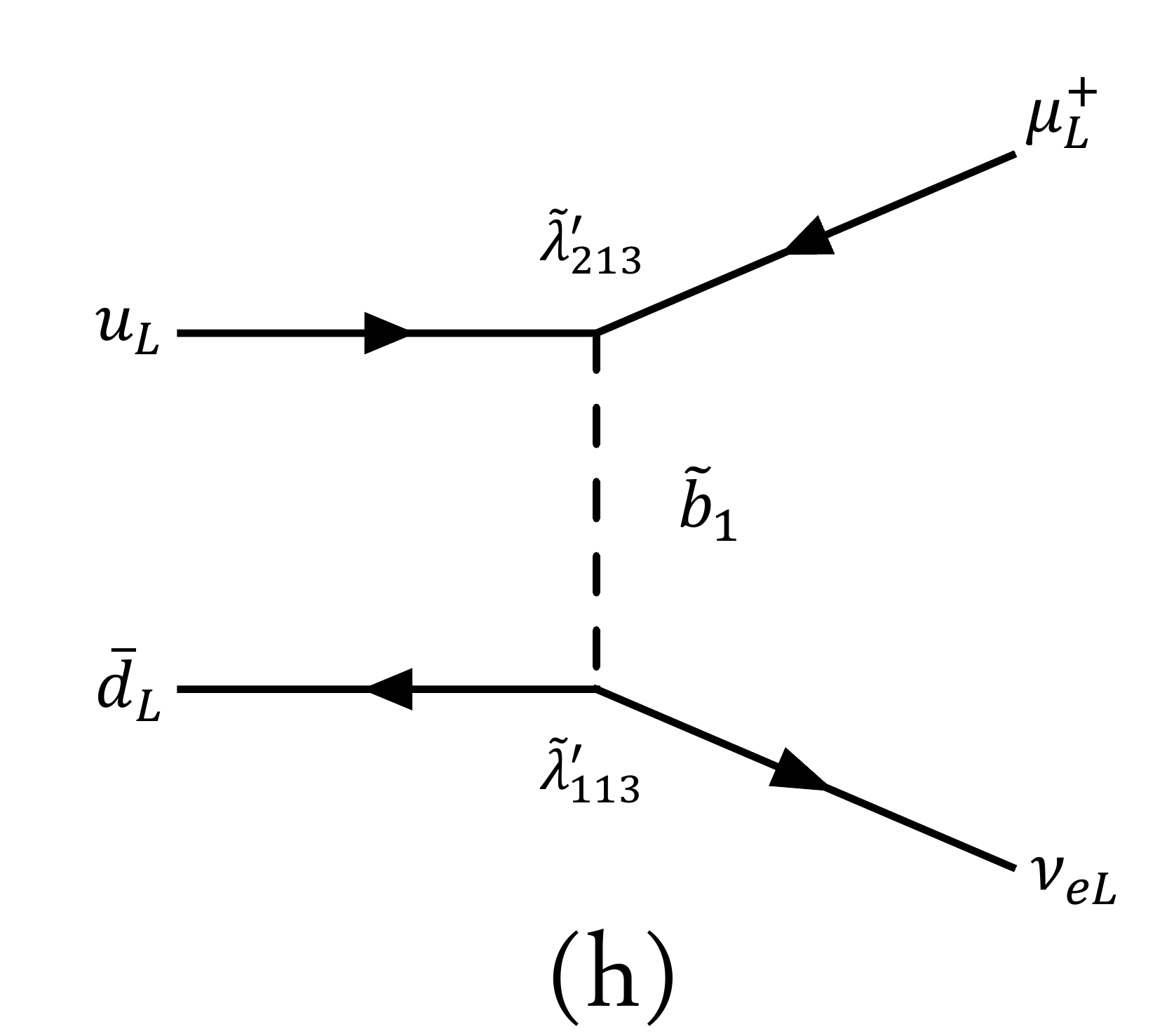}
  \end{minipage}
\caption{RPV contributions to the charged pion decays.}
\label{Fig:pidecaymode2}
\end{figure}

The RPV interactions, Eq.~\eqref{Eq:RPV_L1}, could violate the lepton flavor 
universality of pion decays (Fig.\,\ref{Fig:pidecaymode2}). The RPV 
contributions $\delta\Gamma_e$ and $\delta\Gamma_\mu$ are related 
to the ratio of decay rates as  
\begin{align}
	R^\pi_{e/\mu} 
	= \frac{\Gamma_e^\text{SM} + \delta\Gamma_e}
	{\Gamma_{\mu}^\text{SM} + \delta\Gamma_\mu}
	\simeq \beta \left(1 + \epsilon_e - \epsilon_\mu \right), 
\label{eq:R^pi}
\end{align}
where $\beta=\Gamma_e^\text{SM}/\Gamma_\mu^\text{SM}$, 
$\epsilon_e=\delta\Gamma_e/\Gamma_e^\text{SM}$, and 
$\epsilon_\mu=\delta\Gamma_\mu/\Gamma_\mu^\text{SM}$. 
Here we assumed $\epsilon_e$, $\epsilon_\mu\ll 1$. 
After the straightforward calculation, we obtain
\begin{align}
	\epsilon_e 
	=& 
	\frac{\bigl| \tilde{\lambda}'_{113} \bigr|^2}{V_{ud}^2 G_F^2 m_1^4}
	\left\{
	\frac{\bigl| \tilde{\lambda}'_{231} \bigr|^2 + \bigl| \tilde{\lambda}'_{131} \bigr|^2}{128}
	\Bigl( \frac{m_\pi}{m_u+m_d} \Bigr)^2 
	\frac{m_\pi^2}{m_e^2} 
	+ \frac{V_{ud} G_F m_1^2}{2\sqrt{2}} 
	+ \frac{\bigl| \tilde{\lambda}^{\prime}_{113} \bigr|^2 
	+ \bigl| \tilde{\lambda}^{\prime}_{213} \bigr|^2}{32}
	\right\},
\label{eq:epsilon_e}
\end{align}
\begin{align}
	\epsilon_\mu 
	=& 
	\frac{\bigl|\tilde{\lambda}'_{213} \bigr|^2}{V_{ud}^2 G_F^2 m_1^4} 
	\left\{ 
	\frac{\bigl| \tilde{\lambda}'_{231} \bigr|^2 + \bigl| \tilde{\lambda}'_{131} \bigr|^2}{128} 
	\Bigl( \frac{m_\pi}{m_u+m_d} \Bigr)^2 
	\frac{m_\pi^2}{m_\mu^2} 
	+ \frac{V_{ud} G_F m_1^2}{2\sqrt{2}} 
	 +\frac{\bigl| \tilde{\lambda}^{\prime}_{213} \bigr|^2 
	 +\bigl| \tilde{\lambda}^{\prime}_{113} \bigr|^2}{32}
	 \right\},
\label{eq:epsilon_mu}
\end{align}
Here $V_{ud}$ is the u-d component of the CKM matrix.
The first term in the parenthesis comes from the diagrams (a) and (b) in 
Fig.\,\ref{Fig:pidecaymode2} for 
$\epsilon_e$, and from the diagrams (e) and (f) for $\epsilon_\mu$. 
These initial states form a scalar state with $u_L$ and $d_R$, and their 
contributions are much bigger than other diagram's ones by 
$m_\pi^2/(m_u + m_d)^2$, which is so-called chiral enhancement 
effect\,\cite{Vainshtein:1975sv}. 
In the parameter region we are interested in, the first terms dominate 
$\epsilon_e$ and $\epsilon_\mu$. Besides, the direct search limit 
\eqref{Eq:direct} is more stringent than the limits from the second and 
third terms of Eqs.~\eqref{eq:epsilon_e} and \eqref{eq:epsilon_mu}. 
Then the second and third terms are irrelevant in our analysis. 
The experimental constraint is given by $R^{\pi,\rm{exp}}_{e/\mu}=1.2327(23)\times 10^{-4}$ according to Ref.~\cite{Zyla:2020zbs}.
With the SM prediction $R_{e/\mu}^{\pi,\rm{SM}} = 1.2352 \times 
10^{-4}$\,\cite{Marciano:1993sh, Cirigliano:2007ga}, we set 
the constraint as
\begin{align}
	-7 \times 10^{-7} < 
	\beta \left( \epsilon_e - \epsilon_\mu \right) < 2 \times 10^{-7},
\label{Eq:LFUbound}	
\end{align}
where we allow for a discrepancy of 2$\sigma$.

The RPV interactions also affect the decay $\pi^0\to e^+e^-$. Since the RPV 
interactions lead to a (pseudo-)vector state for the initial state, this decay 
mode does not receive the chiral enhancement. It means that, as long as 
$|\lambda'|^2/m_1^2 \ll G_F$, the RPV effects do not appear on this mode. 
It is because even the $Z^0$ exchange channel is negligible compared with 
leading channel, i.e., the electromagnetic loop one\,\cite{Bergstrom:1982zq}.

\subsection{Neutrinoless double beta decay}
\label{sec:0nu2beta}

\begin{figure}[htbp]
\centering
\includegraphics[keepaspectratio, width=7cm]{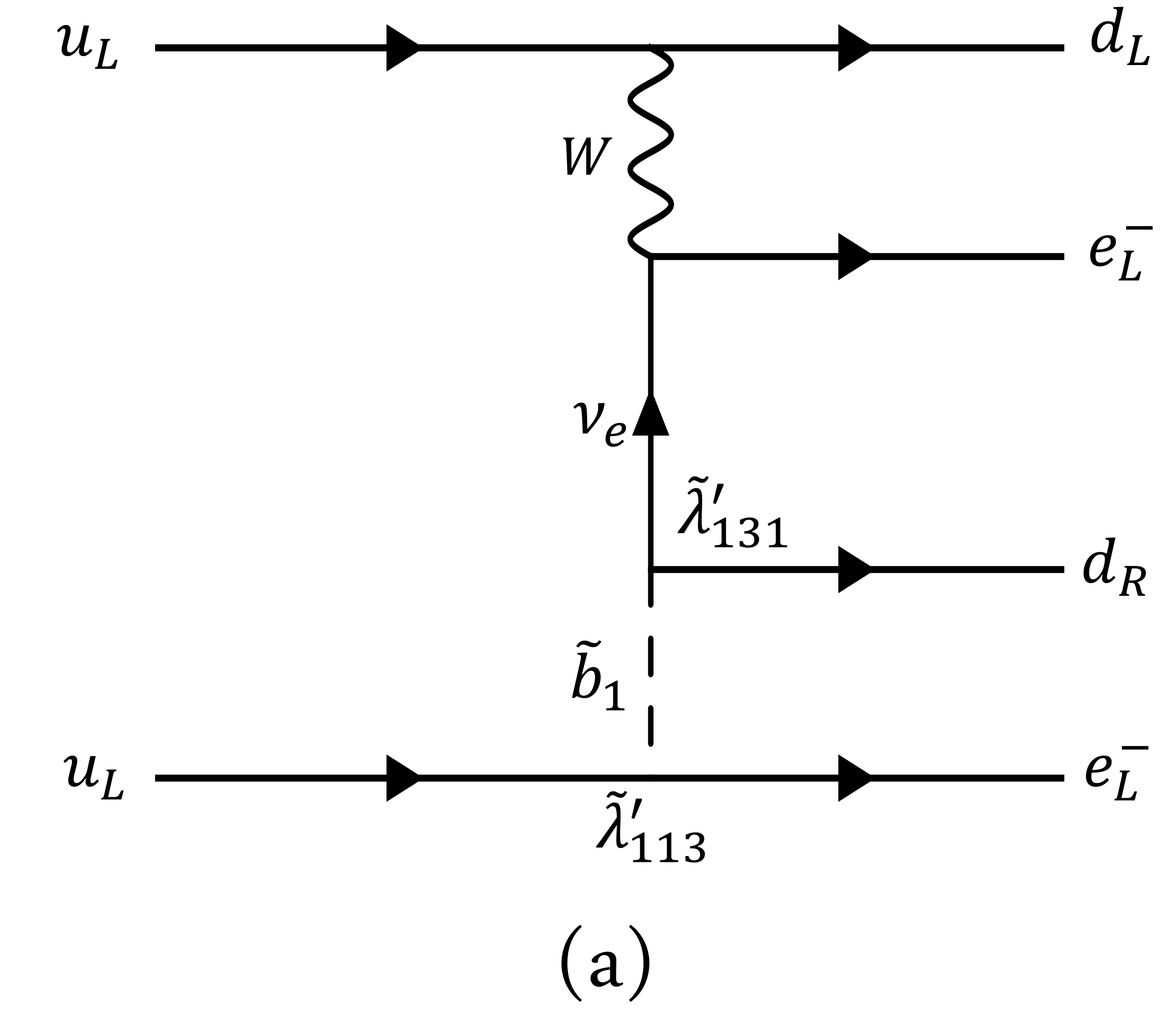}
\includegraphics[keepaspectratio, width=7cm]{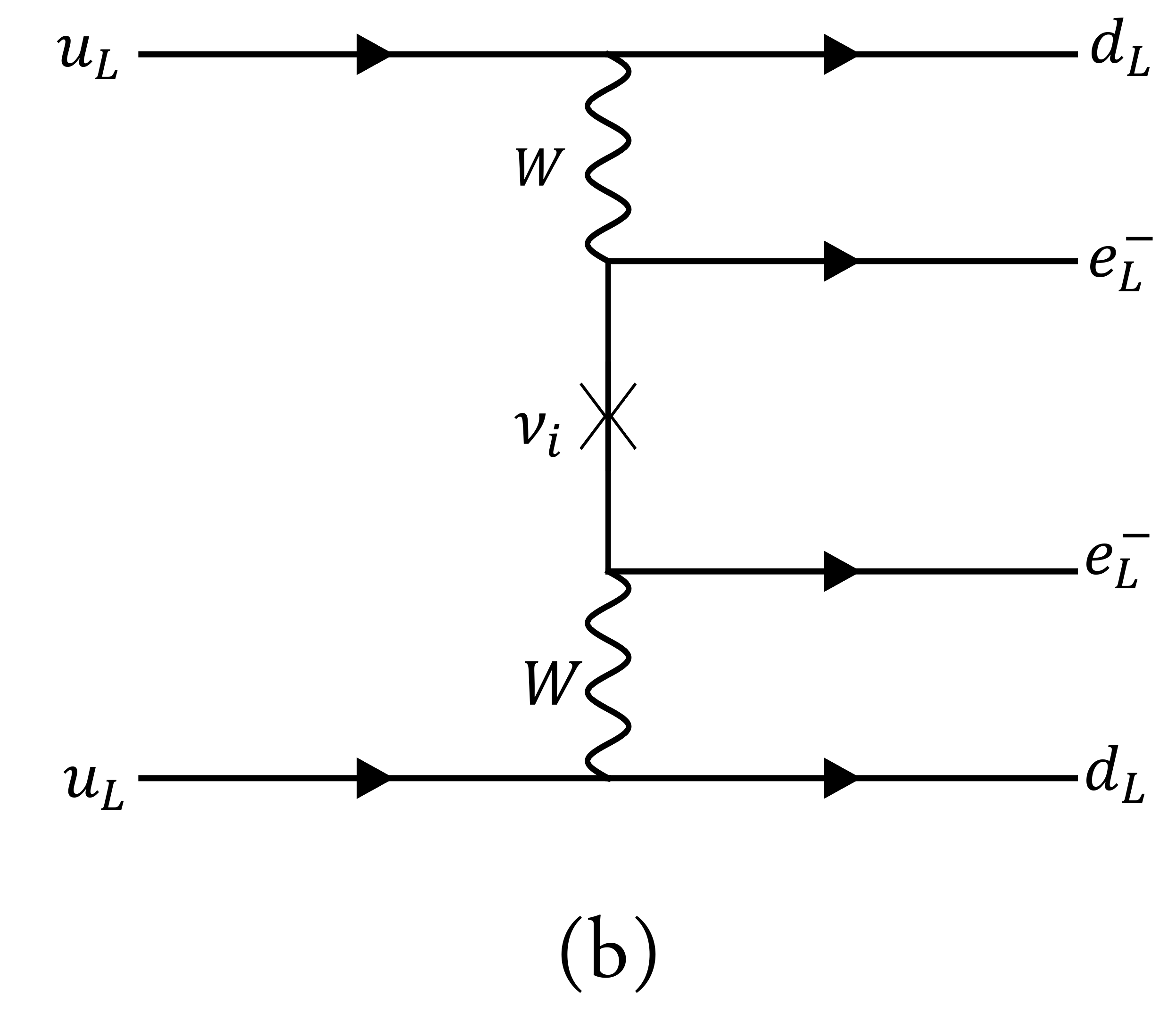}
\caption{$0\nu2\beta$ in the RPV scenario (a) and in the 
Majorana neutrino scenario (b). }
\label{fig:0nu2betadiagram}
\end{figure}

We estimate the bound on RPV parameters along with the 
neutrinoless double beta decay ($0\nu2\beta$) in analogy with that 
assuming the Majorana neutrinos. 
Extracting the LNV source part in each amplitude (Fig.\,\ref{fig:0nu2betadiagram}), 
we find the relation 
\begin{align}
	\frac{\bigl| \tilde{\lambda}'_{131} \tilde{\lambda}'_{113} \bigr|}{2m_1^2} q 
	&\simeq \frac{4V_{ud} G_F}{\sqrt{2}} \overline{m_{ee}},
\label{eq:0nu2beta}
\end{align}
where $\overline{m_{ee}}$ is the effective Majorana mass of electron 
neutrino and $q$ is the momentum of internal neutrino. In our analysis 
we set $q=100$\,MeV, which is evaluated by the typical distance 
between nucleons in a nucleus. Applying the bound 
$\overline{m_{ee}} \lesssim 0.1$\,eV\,\cite{Zyla:2020zbs}, above 
relation \eqref{eq:0nu2beta} leads to the bound on RPV parameters as 
\begin{align}
	\bigl| \tilde{\lambda}'_{131} \tilde{\lambda}'_{113} \bigr| 
	< 2.6\times10^{-9} \left( \frac{m_1}{200\mbox{\rm (GeV)}} \right)^2.
\end{align}

\section{New physics searches using muonic atoms}
\label{Sec:muonic_atom}

The muonic atom sheds light on not only the LFV but also the LNV through 
$\mu^-\to e^-$ conversion and $\mu^-\to e^+$ conversion.

\subsection{$\mu^- \to e^-$ conversion}
\label{sec:m2e-}

The $\mu^- \to e^-$ conversion in nuclei occurs with the combination of 
$\tilde{\lambda}'_{213}$ and $\tilde{\lambda}'_{113}$ 
(Fig.~\ref{fig:diagram_mu2e-}). Applying the formula for $\mu^- \to e^-$ 
conversion rate\,\cite{Kitano:2002mt}, the branching ratio in our scenario 
is obtained by 
\begin{align}
	B \left(\mu^- \to e^-; N \right) 
	= \tilde{\tau}_\mu 
	\frac{\bigl| \tilde{\lambda}'_{213} \tilde{\lambda}'_{113} \bigr|^2}{4m_1^4} 
	\left(2V^{(p)} + V^{(n)} \right)^2 m_\mu^5.
\end{align}
The dimensionless overlap integral $V^{(p,n)}$ and the muonic-atom lifetime 
$\tilde{\tau}_\mu$ are listed in Table~\ref{tab:mueconpara1}. 
The most stringent bound, $B \left(\mu^- \to e^-; \mathrm{Au} \right) < 
7 \times 10^{-13}$\,\cite{Bertl:2006up}, gives the limit by 
$\bigl| \tilde{\lambda}'_{213} \tilde{\lambda}'_{113} \bigr| < 1.6 \times 
10^{-7}$ for $m_1=200$\,GeV.

\begin{table}[htb] 
\caption{Overlap integrals $V^{(p)}$ and $V^{(n)}$~\cite{Kitano:2002mt} and the 
lifetime of a muonic atom \cite{Suzuki:1987jf}.}
\begin{tabular}{cccc}\hline
	Nucleus & $V^{(p)}$ &  $V^{(n)}$ & $\tilde{\tau}_\mu$ [ns] \\ \hline
	$^{27}$Al & $0.0161$ & $0.0173$ & 864.0 \\ 
	$^{197}$Au & $0.0974$ & $0.146$ & 74.3 \\ \hline
\end{tabular}
\label{tab:mueconpara1}
\end{table}

\subsection{$\mu^- \to e^+$ conversion}
\label{Sec:mutoposi}

The combination of the LFV RPV couplings and the $\tilde{b}_L$-$\tilde{b}_R$ 
mixing gives rise to the $\mu^-\to e^+$ conversion in nuclei (Fig.~\ref{fig:RPV}).
The reaction rate of $\mu^-\to e^+$ conversion faces the nuclear transition matrix. 
For the Majorana-neutrino case, it is evaluated by the nuclear proton-neutron 
renormalized quasi-particle random phase approximation \cite{Simkovic:2000ma, 
Vergados:2002pv,Domin:2004tk} and the shell model calculation \cite{Divari:2002sq}. 
The short-range effective operators inducing the $\mu^- \to e^+$ conversion were 
discussed in Refs.~\cite{Geib:2016daa,Geib:2016atx,Berryman:2016slh}. 
The conversion rate for other types of operators have not been qualitatively investigated, 
also for the operator in this work. We therefore estimate the conversion rate in analogy 
with the muon capture $\mu^-p\to \nu_\mu n$ in muonic atoms.

We adopt the phenomenological parametrization of capture rate for a nucleus of an 
atomic number $Z$ and of a mass number $A$~\cite{Primakoff:1959fs, Suzuki:1987jf}, 
\begin{align}
	\Gamma_\textrm{cap} \simeq Z_\mathrm{eff}^4 X_1 
	\left(1-X_2\frac{A-Z}{2A}\right). 
\label{eq:capture_rate1}
\end{align}
Here $Z_\mathrm{eff}$ is the effective atomic number for muonic atoms~\cite{FORD1962295}.
The $Z_\mathrm{eff}$ dependence stems from the effective number of protons in a nucleus ($Z_\mathrm{eff}$) and probability of a muon being at the nuclear center ($Z_\mathrm{eff}^3$); the latter can also be understood by the expression of the muon wave function, $\left|\psi_\mu(0)\right|^2=\left(m_\mu Z_\mathrm{eff}\alpha\right)^3/\pi$.
The parameter $X_1$ corresponds to the capture rate for muonic hydrogen, and $X_2$ parametrizes the Pauli blocking effect.
The experimental data fit the parameters by $X_1=170$\,s$^{-1}$ and $X_2=3.125$~\cite{Suzuki:1987jf}.

The $\mu^- \to e^+$ conversion rate is inferred in an analogy of the muon 
capture rate as
\begin{equation}
\begin{split}
	\Gamma (\mu^- \to e^+; N) 
	\simeq 
	\frac{\bigl| \tilde{\lambda}^\prime_{2ij} \tilde{\lambda}^\prime_{1ji} \bigr|^2}{m_1^4}
	\left( \frac{G_F}{\sqrt{2}} \right)^2 
	\frac{1}{q^2}Q_{\mu^-\to e^+}^{8} 
	Z_\mathrm{eff}^2  \left| \psi_\mu(0) \right|^2 
	\left( 1-X_2\frac{A-Z}{2A} \right)^2,
\label{Eq:Ratemu-toe+}	
\end{split}
\end{equation}
where $N$ presents the initial nucleus, and $\left( i, j \right) = \left( 1, 3 \right), \left( 3, 1 \right)$. 
The factor 
$1/q^2$ expresses the correlation function of active neutrino of momentum 
$q$. Since the process associates with the internal conversion $2p \to 2n$, 
it is expected that the rate is proportional to $Z_\mathrm{eff}^2$. Note that 
the energy scale factor $Q_{\mu^-\to e^+}$ contains the nuclear transition 
strength in addition to the phase space volume, and its power is determined 
by the dimensional analysis. 
The branching ratio is given by $B\left( \mu^- \to e^+ ; N \right) = 
\tilde{\tau}_\mu \Gamma \left( \mu^- \to e^+ ; N \right)$. 
We use $X_2=3.125$ as the muon capture, and we take $q= 100$\,MeV, 
which corresponds to the Fermi momentum of nucleon in the nucleus. 
The energy scale factor is set by $Q_{\mu^-\to e^+} = m_\mu$.

The current experimental bound is $B\left( \mu^- \to e^+; \text{Ti} \right) 
< 1.7 \times 10^{-12}$ ($3.6\times 10^{-11}$) for the transition to the ground 
(giant dipole resonance) state of calcium~\cite{SINDRUMII:1998mwd}.
Using Eq.~\eqref{Eq:Ratemu-toe+}, we obtain the limit by $\left|\tilde{\lambda}'_{2ij}\tilde{\lambda}'_{1ji}\right|^2/m_1^4 < 6.5\times 10^{-21}$ MeV$^{-4}$.

\section{Results}
\label{Sec:Result}

Numerical analysis is shown in two cases; One is of negligible 
$\mu^-\to e^-$ conversion rate and the other is more general ones. 
We adopt $m_1 = 200$\,GeV. 
Free parameters are the four RPV couplings, $\tilde{\lambda}'_{213}$, 
$\tilde{\lambda}'_{131}$, $\tilde{\lambda}'_{113}$, and 
$\tilde{\lambda}'_{231}$.

\subsection{Case of no $\mu^-\to e^-$ conversion} 
\label{sec:without_mu2e-}

We separately investigate two patterns wherein the $\mu^-\to e^-$ 
conversion is turned off: 
(pattern I)  $\tilde{\lambda}'_{213} \neq 0$, 
$\tilde{\lambda}'_{131} \neq 0$, 
and $\tilde{\lambda}'_{113} = \tilde{\lambda}'_{231} = 0$ ~~ 
(pattern II) $\tilde{\lambda}'_{113}\neq 0 $, $\tilde{\lambda}'_{231} \neq 0$, 
and $\tilde{\lambda}'_{213} = \tilde{\lambda}'_{131} = 0$.

\begin{figure}[htbp]
\centering
\begin{minipage}[b]{0.49\linewidth}
\centering
\includegraphics[width=0.8\linewidth]{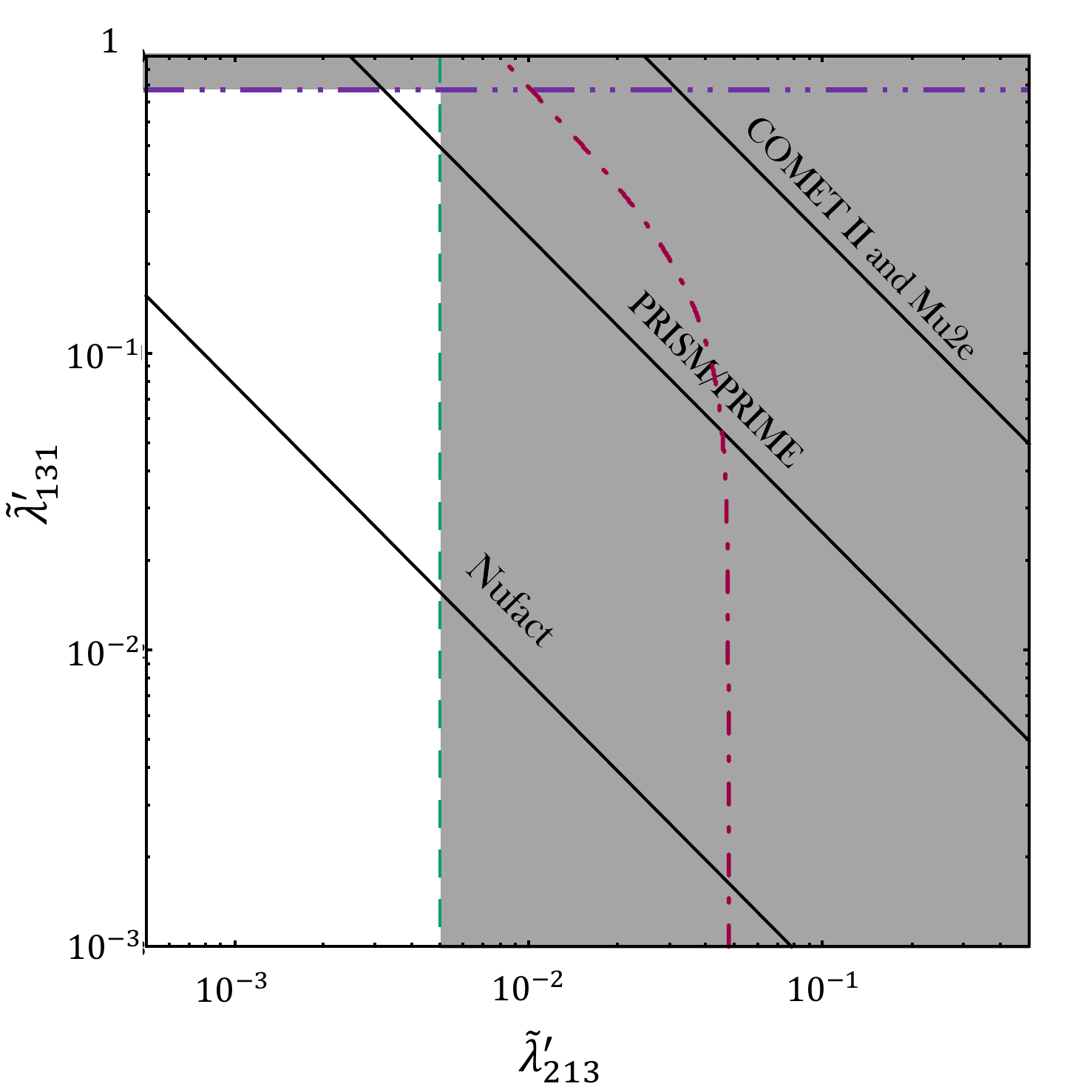}
\end{minipage}
\begin{minipage}[b]{0.49\linewidth}
\centering
\includegraphics[width=0.8\linewidth]{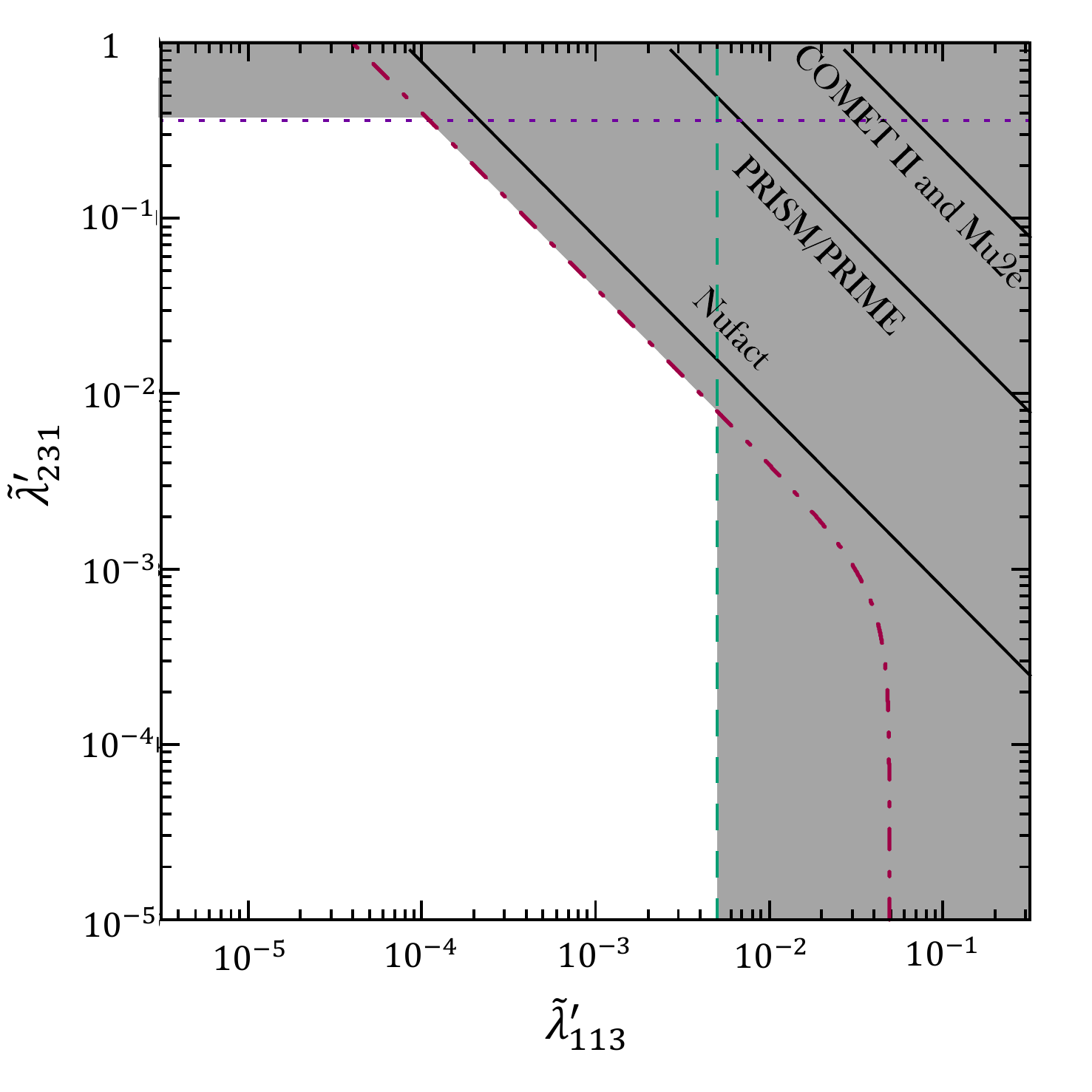}
\end{minipage}
\caption{$B \left(\mu^- \to e^+; \text{Ca} \right) = 10^{-16}$, 
$10^{-18}$, and $10^{-21}$ (black solid) corresponding to the muon 
productions at COMET phase-II (Mu2e), PRISM/PRIME, and Nufact 
experiments in the Pattern I (left) and Pattern II (right). 
The excluded regions (shaded area) comprise the bounds from 
the direct sbottom search (dashed green), the LFU in charged pion 
decays (dash-dotted red), the APV and PVES for 
$m_{\tilde{t}_L} =1$~TeV (dashed-two-dotted purple), and 
$\nu_\mu d_R\to \nu_\mu d_R$ (dotted purple).}
\label{fig:Ratecoupling}
\end{figure}

\subsubsection{Pattern I: $\tilde{\lambda}'_{213} \neq 0$, 
$\tilde{\lambda}'_{131} \neq 0$, 
and $\tilde{\lambda}'_{113} = \tilde{\lambda}'_{231} = 0$} 
\label{sec:patternI}

We evaluate the maximal $B\left(\mu^- \to e^+; N\right)$. 
Decomposing $B\left(\mu^- \to e^+; N\right)$ into the target 
dependent part $\tilde{B}$ (Table~\ref{tab:Btilde}) and uncertain 
parts ($q$ and $Q_{\mu^-\to e^+}$), it is rewritten as
\begin{align}
	B\left(\mu^- \to e^+; N \right) 
	= 
	\tilde{B} 
	\left( \frac{100\,(\mbox{\rm MeV})}{q} \right)^2 
	\left( \frac{Q_{\mu^-\to e^+}}{100\,(\mbox{\rm MeV})} \right)^8. 
\end{align}
We find $B\left(\mu^- \to e^+; N \right) \sim \mathcal{O} 
(10^{-18})$ in the pattern I. 
The COMET phase-II (Mu2e), PRISM/PRIME, and Nufact experiments 
respectively plan to accumulate $\mathcal{O} \left(10^{16}\right)$, 
$\mathcal{O} \left(10^{18}\right)$, and $\mathcal{O} 
\left(10^{21}\right)$ muons.  Figure~\ref{fig:Ratecoupling} (a) 
shows $B \left( \mu^- \to e^+; \text{Ca} \right) = 10^{-16}$, 
$10^{-18}$, and $10^{-21}$ (black solid) corresponding to these 
muon productions. The calcium (Ca) target would maximize the 
$S/N$ ratio\,\cite{Yeo:2017fej}. 
The shaded area shows the excluded parameter region. 
$\tilde{\lambda}'_{131}$ is unbound unless the stop mass is 
given. Here we take $m_{\tilde{t}_L}=1$~TeV. Then the direct 
search~\eqref{Eq:direct} and the measurement of 
APV-PVES~\eqref{Eq:bound-131} draw the boundaries. The 
bound on $\tilde{\lambda}'_{131}$ gets looser for the heavier 
stop mass, and then the LFU test in pion decays makes the boundary. 
It is testable in near future experiments, and could shed light on 
the LFV and LNV sources.

It is important to emphasize that the LFU in pion decays tightly 
correlated with the $\mu^-\to e^+$ conversion (compare 
Figs.~\ref{fig:RPV} and \ref{Fig:pidecaymode2}) in this scenario. 
When the violation of LFU is observed in pion decays, searches for 
the $\mu^-\to e^+$ conversion would provide complementary 
information for new physics.

\begin{table}[htb]
\centering
\caption{Target dependent coefficient $\tilde{B}$ by 
$(\tilde{\lambda}'_{131}, \tilde{\lambda}'_{213}) = (6.9\times 10^{-1}, 5.0\times 10^{-3})$ 
for the pattern I and $(\tilde{\lambda}'_{113}, \tilde{\lambda}'_{231}) 
= (5.0\times 10^{-3}, 8.0\times 10^{-3})$ for the pattern II which leads to 
the maximal $B\left(\mu^- \to e^+; N \right)$,  
effective atomic number $Z_\mathrm{eff}$, and lifetime 
of the muonic atom $\tilde{\tau}_\mu$\,\cite{Suzuki:1987jf}.}
\vspace{0.4cm}
\doublerulesep 0.8pt \tabcolsep 0.4cm
\begin{tabular}{ccccc} \hline\hline
	Nucleus & $Z_\mathrm{eff}$ & $\tilde{\tau}_\mu$ [ns] & $\tilde{B}$ (Pattern I) & $\tilde{B}$ (Pattern II) \\ \hline
	$^{27}$Al & 11.48 & 864 & $7.0\times 10^{-19}$ & $9.2\times 10^{-23}$ \\
	$^{32}$S & 13.64 & 540 & $1.4\times 10^{-18}$ & $1.8\times 10^{-22}$ \\
	$^{40}$Ca & 16.15 & 333 & $2.0\times 10^{-18}$ & $2.6\times 10^{-22}$ \\
	$^{48}$Ti & 17.38 & 330 & $1.4\times 10^{-18}$ & $1.8\times 10^{-22}$ \\
	$^{65}$Zn & 21.61 & 161 & $2.2\times 10^{-18}$ & $2.8\times 10^{-22}$ \\
	$^{73}$Ge & 22.43 & 167.4 & $1.6\times 10^{-18}$ & $2.1\times 10^{-22}$ \\\hline\hline
\end{tabular}
\label{tab:Btilde}
\end{table}

\subsubsection{Pattern II: $\tilde{\lambda}'_{113}\neq 0 $, 
$\tilde{\lambda}'_{231} \neq 0$, 
and $\tilde{\lambda}'_{213} = \tilde{\lambda}'_{131} = 0$} 
\label{sec:patternII}

Applying $\tilde{B}$ in Table~\ref{tab:Btilde} for the pattern II, the 
maximal $B\left(\mu^- \to e^+; N\right)$ is obtained by $\mathcal{O} 
(10^{-21})$. This implies that the discovery of $\mu^- \to e^+$ 
conversion at COMET, Mu2e, and PRISM experiments rules out the 
pattern II. 
Figure~\ref{fig:Ratecoupling} (b) is the same as Fig.~\ref{fig:Ratecoupling} 
(a) but for the $\tilde{\lambda}'_{113}$-$\tilde{\lambda}'_{231}$ plane. 
The direct search~\eqref{Eq:direct} and the LFU test in pion 
decays~\eqref{Eq:LFUbound} draw the boundaries to the excluded region.

\subsection{General analysis including all four couplings} 
\label{sec:all_couplings}

The bounds on RPV couplings from the searches for $\mu^-\to e^-$ 
conversion and $0\nu2\beta$ are also comprehended, in addition to the 
bounds discussed in Sec.\,\ref{sec:without_mu2e-}. 
The bounds derived from relevant observables are 
summarized in Table~\ref{tab:boundonallcoupling}.

\begin{center}
\begin{table}[htb]
\caption{Bounds on the RPV couplings applied in 
Sec.~\ref{sec:all_couplings}. Here $m_1 = 200$\,GeV. 
$\beta = \Gamma_e^{\text{SM}}/\Gamma_\mu^{\text{SM}}$, 
$\epsilon_e$, and $\epsilon_\mu$ are given in 
Eqs.~\eqref{eq:epsilon_e} and \eqref{eq:epsilon_mu}.}
\begin{center}
\begin{tabular}{ccc} \hline
Observables & Bound & Section 
\\ \hline\hline
APV and PVES & $\tilde{\lambda}'_{131}\leq 0.69$ & \ref{sec:APV_PVES} 
\\ \hline 
$\nu_\mu d_R \to \nu_\mu d_R$ 
& $\tilde{\lambda}'_{231}\leq 0.36$ 
& \ref{sec:DIS} 
\\ \hline 
Direct sbottom search  
& \ $\tilde{\lambda}'_{i13}\leq 5\times 10^{-3}$ ($i=1,2$) 
& \ref{sec:sbottom_decay}  
\\ \hline 
LFU of $\pi^{\pm}$ decays 
& $-7 \times 10^{-7} \leq \beta \left( \epsilon_e - \epsilon_\mu \right) \leq 2 \times 10^{-7}$
& \ref{sec:lepton_universality} 
\\ \hline 
$0\nu2\beta$ 
& $\tilde{\lambda}'_{113}\tilde{\lambda}'_{131}\leq2.6\times10^{-9}$ 
& \ref{sec:0nu2beta} 
\\ \hline 
$\mu^-\to e^-$ conversion 
& $\tilde{\lambda}'_{213}\tilde{\lambda}'_{113}\leq1.6\times10^{-7}$ 
& \ref{sec:m2e-} 
\\ \hline 
\end{tabular}
\label{tab:boundonallcoupling}
\end{center}
\end{table}
\end{center}

\begin{figure}[ht]
\begin{minipage}[b]{0.49\linewidth}
\centering
\includegraphics[width=0.8\linewidth]{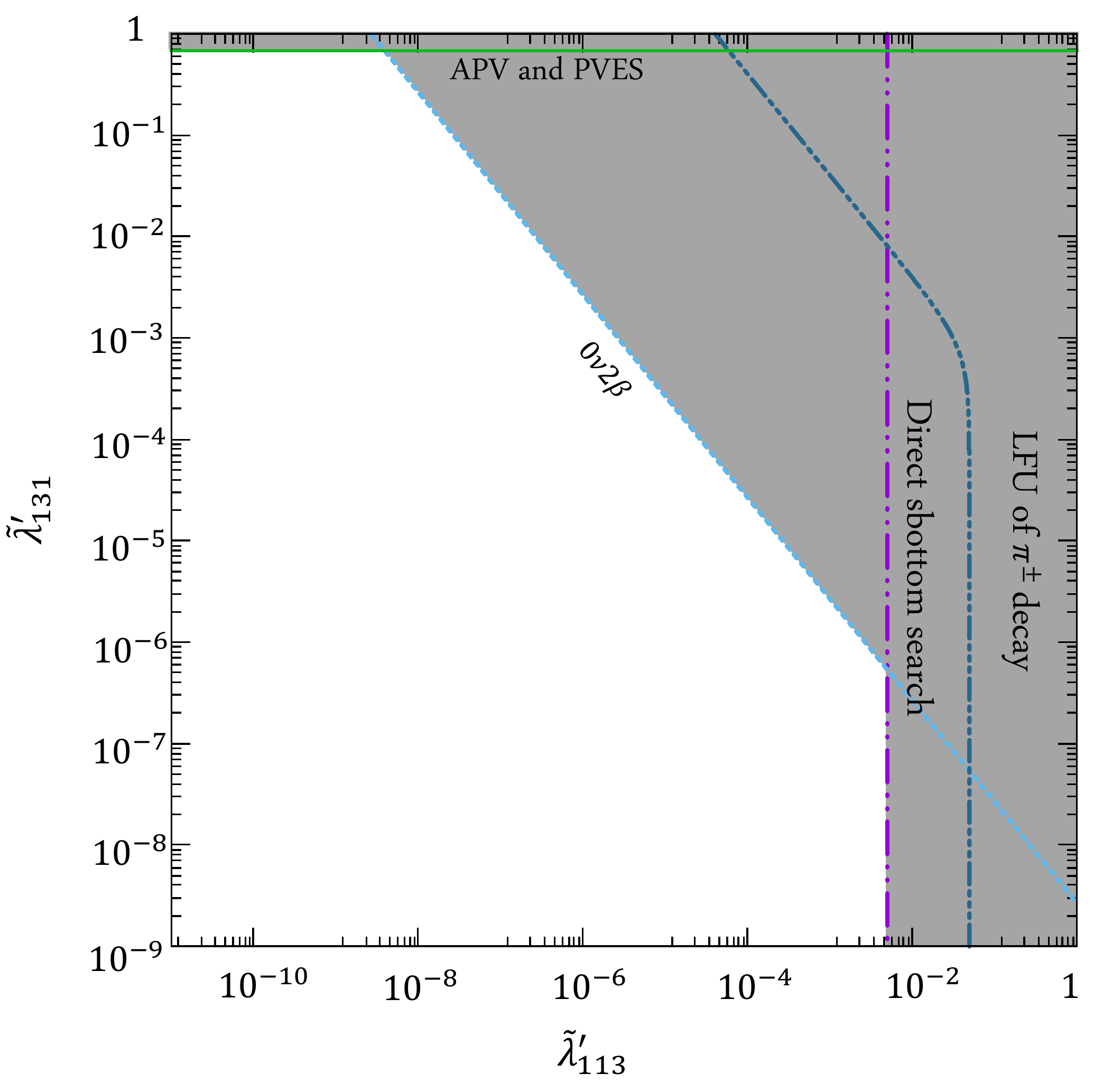}
\end{minipage}
\begin{minipage}[b]{0.49\linewidth}
\centering
\includegraphics[width=0.8\linewidth]{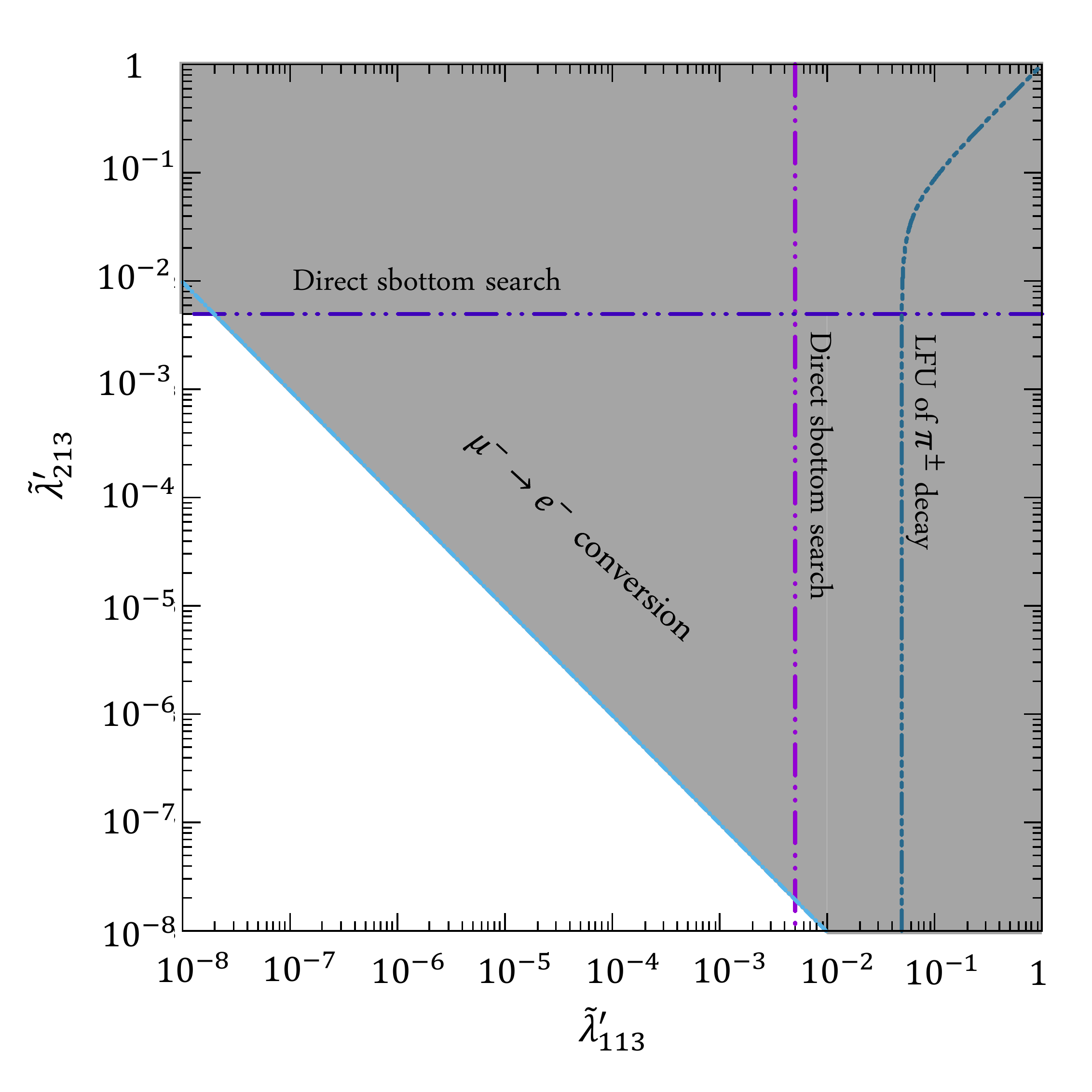}
\end{minipage}
\begin{minipage}[b]{0.49\linewidth}
\centering
\includegraphics[width=0.8\linewidth]{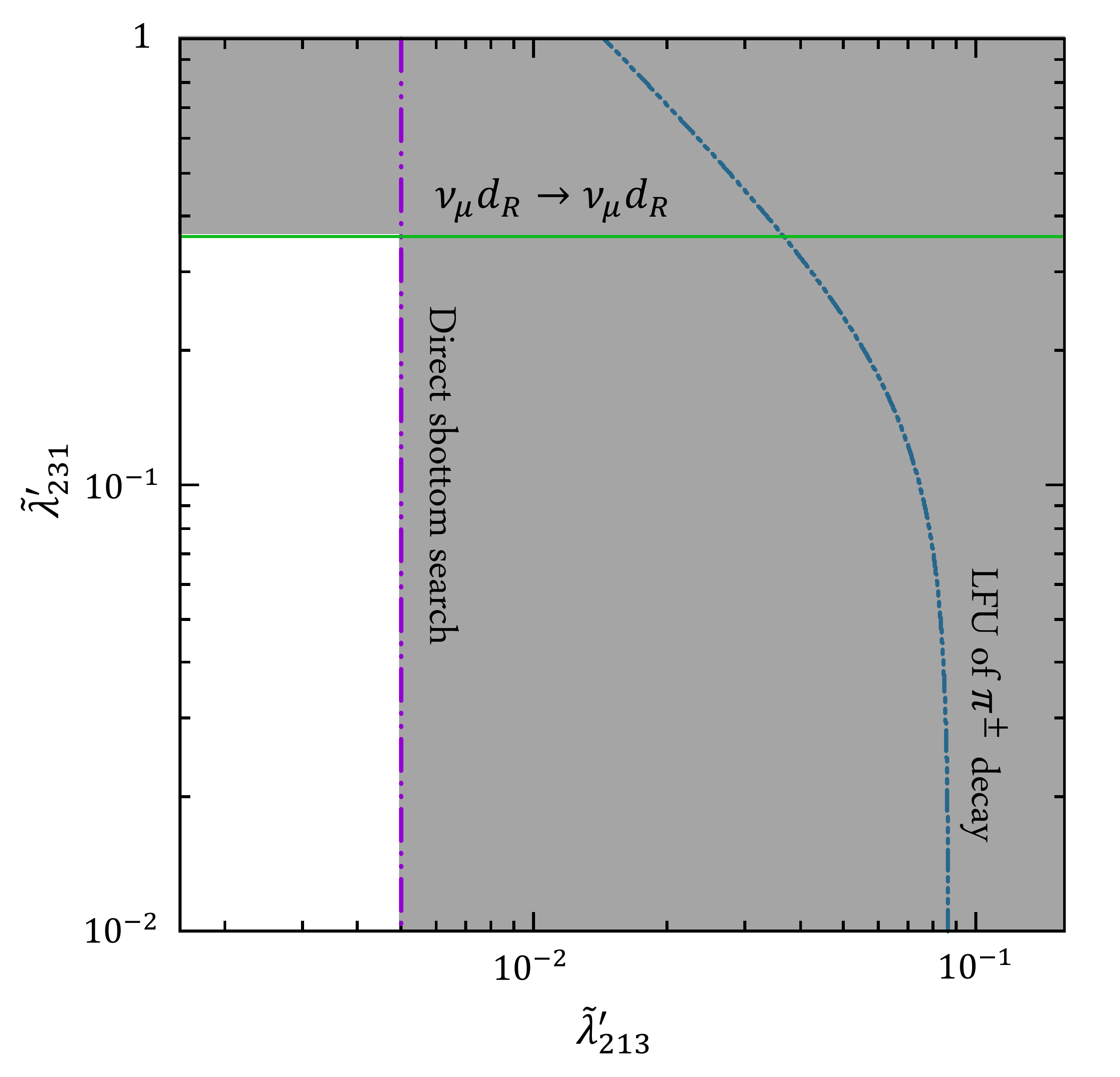}
\end{minipage}
\begin{minipage}[b]{0.49\linewidth}
\centering
\includegraphics[width=0.8\linewidth]{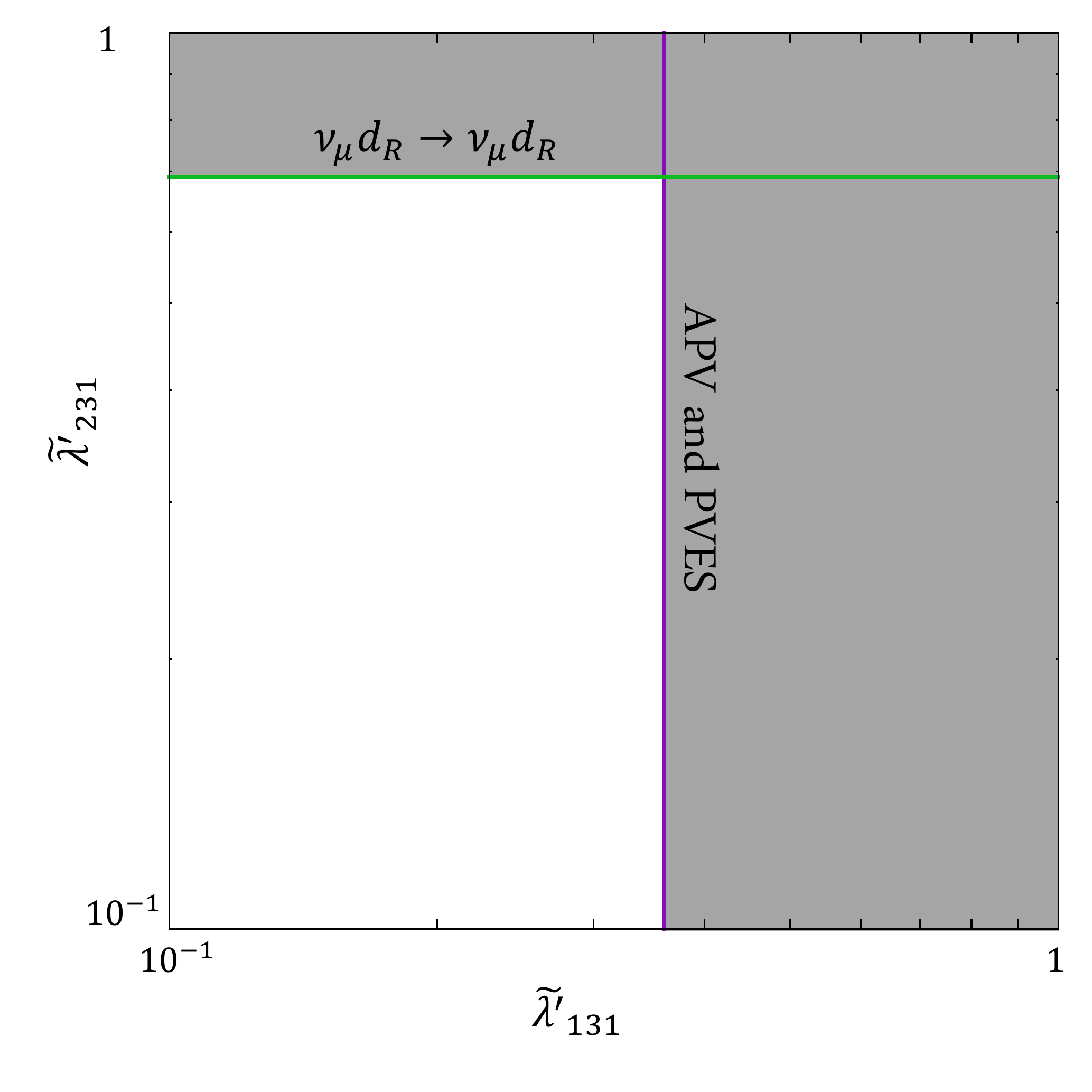}
\end{minipage}
\caption{Excluded regions (shaded area) in the plane of two couplings. 
The observables attached on each line draw the boundaries according to 
Table~\ref{tab:boundonallcoupling}. In the investigation for a combination 
of two RPV couplings, other RPV couplings are set to be zero.}
\label{fig:relation_couplings}
\end{figure}

Figure~\ref{fig:relation_couplings} shows the excluded region (shaded 
area) for each combination of RPV couplings. 
In each panel, the other RPV couplings are set to be zero. 
It has been already investigated for the 
$\tilde{\lambda}'_{131}$-$\tilde{\lambda}'_{213}$ and 
$\tilde{\lambda}'_{113}$-$\tilde{\lambda}'_{231}$ planes, wherein 
both the $\mu^-\to e^-$ conversion and $0\nu 2\beta$ are turned off. 
The $0\nu 2\beta$ search draws the outline of excluded region in the 
$\tilde{\lambda}'_{113}$-$\tilde{\lambda}'_{131}$ plane 
(Fig.\,\ref{tab:boundonallcoupling} (a)). The $\mu^-\to e^-$ 
conversion search draws the outline of excluded region in the 
$\tilde{\lambda}'_{113}$-$\tilde{\lambda}'_{213}$ plane 
(Fig.\,\ref{tab:boundonallcoupling} (b)). 
These processes are therefore important ingredients for the analysis in 
the space of $\bigl( \tilde{\lambda}'_{113}$, $\tilde{\lambda}'_{131}$, 
$\tilde{\lambda}'_{213}$, $\tilde{\lambda}'_{231} \bigr)$.

\begin{figure}[ht]
\centering
\includegraphics[keepaspectratio, width=12cm]{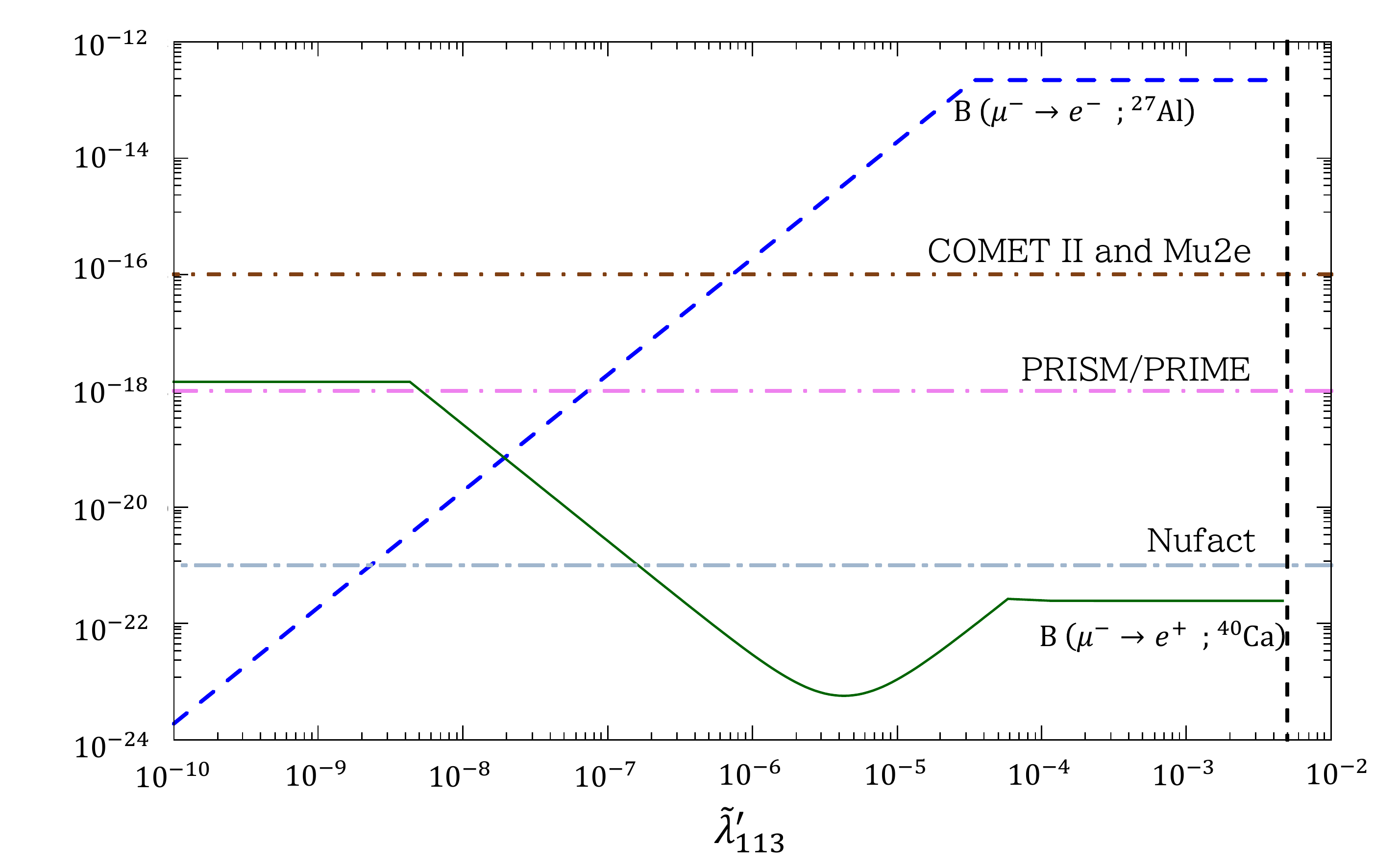}
\caption{$\lambda'_{113}$ dependence of maximally 
allowed $B\left(\mu^-\to e^+; \text{Ca}\right)$ and 
$B\left(\mu^-\to e^-; \text{Al}\right)$. Inverse of 
expected muon productions at each experiment are 
shown by horizontal lines, $B=10^{-16}$ (COMET phase-II and Mu2e), 
$10^{-18}$  (PRISM/PRIME), and $10^{-21}$ (Nufact). 
The direct search bound $\tilde{\lambda}'_{113} = 
5 \times 10^{-3}$ is shown by the vertical dashed line.} 
\label{fig:mueVSmutoposi113}
\end{figure}

The bounds and observables in Fig.~\ref{fig:relation_couplings} have 
actually more complicated correlations with each other. 
Figure~\ref{fig:mueVSmutoposi113} shows an example result. 
In the parameter space wherein free from all experimental bounds 
except for the $\mu^- \to e^-$ conversion, first, we lead the 
maximally allowed $B\left(\mu^- \to e^+; \text{Ca}\right)$. 
With these arrangements, the maximally allowed 
$B\left(\mu^-\to e^-; \text{Al} \right)$ is evaluated. 
For the region of $\tilde{\lambda}'_{113} \lesssim 4 \times 10^{-9}$, 
as is close to the setting in Sec.\,\ref{sec:patternI}, 
$B\left(\mu^-\to e^-; \text{Al} \right)$ does not reach the 
PRISM/PRIME sensitivity. In this region, the maximized combination 
$\tilde{\lambda}'_{131}\tilde{\lambda}'_{213}$ leads to the large 
$B\left(\mu^-\to e^+; \text{Ca}\right)$. 
For the region of $4 \times 10^{-9} \lesssim \tilde{\lambda}'_{113} 
\lesssim 5 \times 10^{-6}$, since the search for $0 \nu 2 \beta$ limits 
the combination $\tilde{\lambda}'_{113}\tilde{\lambda}'_{131}$, 
$B\left(\mu^-\to e^+; \text{Ca}\right)$ decreases with 
$\tilde{\lambda}'_{113}$.  
For the region of $\tilde{\lambda}'_{113} \gtrsim 5 \times 10^{-6}$, 
the $\tilde{\lambda}'_{113} \tilde{\lambda}'_{231}$ term dominates 
over the $\tilde{\lambda}'_{131} \tilde{\lambda}'_{213}$ term in Eq.~\eqref{Eq:Ratemu-toe+}, and $B\left(\mu^-\to e^+; \text{Ca}\right)$ 
increases with $\tilde{\lambda}'_{113}$. 
For the region of $\tilde{\lambda}'_{113} \gtrsim 4 \times 10^{-5}$, 
the measurement for of LFU limits the combination 
$\tilde{\lambda}'_{113} \tilde{\lambda}'_{231}$ (see 
Fig.\,\ref{fig:Ratecoupling}), and $B\left(\mu^-\to e^+; \text{Ca}\right)$ 
levels off at $\simeq 10^{-22}$.

The largest $B\left(\mu^-\to e^+; \text{Ca}\right)$ is achieved for 
$B\left(\mu^-\to e^+; \text{Ca}\right) \lesssim 10^{-20}$. 
Both $\mu^-\to e^-$ and $\mu^-\to e^+$ conversions could be observed 
in near future experiments. Complementary measurements of 
these conversions shed light on not only the LFV source but also the origin 
of LNV in new physics scenarios.

\section{Summary}
\label{sec:Summary}

We have investigated the possibility that the LNV process $\mu^-\to e^+$ 
conversion is observed prior to the LFV process $\mu^-\to e^-$ 
conversion. 
For a reference scenario of our interest, we have focused on RPV SUSY 
models wherein the $SU(2)_L$ doublet and singlet sbottom 
($\tilde{b}_L$ and $\tilde{b}_R$) mixes each other.

When the conservation of lepton flavors is violated by the RPV interactions, they give rise to the $\mu^- \to e^-$ conversion. 
The $\tilde{b}_L$-$\tilde{b}_R$ mixing flips the lepton number on 
the internal sbottom line, and hence the lepton number is no longer 
conserved. The $\mu^- \to e^+$ conversion arises via the LFV vertex 
and the $\tilde{b}_L$-$\tilde{b}_R$ mixing. 
It is important to emphasize that, when either $\lambda'_{213}$ or 
$\lambda'_{113}$ is zero, the $\mu^-\to e^-$ conversion rate goes 
to zero, but the $\mu^-\to e^+$ conversion still could be observable.

We have evaluated the rate of $\mu^-\to e^+$ mediated by the sbottom 
in analogy of the muon capture process in muonic atom. 
Then we have investigated how could the $\mu^-\to e^+$ rate be large 
under the experimental bounds on RPV parameters. 
Bounds come from the $\mu^-\to e^-$ conversion search, the measurement 
of LFU in pion decays, the direct sbottom search at the LHC, and so on. 
Especially, we have found that the LFU in pion decays provides the direct constraints for the $\mu^-\to e^+$ rate because they are connected through the same combinations of the couplings.
The largest $B\left(\mu^-\to e^+; \text{Ca}\right)$ is achieved in the 
parameter region of small $B\left(\mu^-\to e^-; \text{Al}\right)$. 
In some parameter regions, both $B\left(\mu^-\to e^+; \text{Ca}\right)$ 
and $B\left(\mu^-\to e^-; \text{Al}\right)$ are experimentally reachable
at next-generation experiments. Complementary measurements of 
these conversions shed light on not only the LFV source and also the origin 
of LNV in new physics scenarios. It is important to search for and analyze 
the non-standard reactions of muonic atoms without prejudice that the 
LFV reactions always are leading compared with the LNV ones.

\section*{Acknowledgements}

We would like to thank T.~Goto, T.~Kitahara, J.~Kriewald, and A.~M.~Teixeira, 
for fruitful comments.This work is supported in part by the JSPS Grant-in-Aid for
Scientific Research Numbers JP18H01210 (J.S. and Y.U.), JP21H00081 (Y.U.), 
and JP20H05852 (M.Y.), and MEXT KAKENHI Grant Number JP18H05543 (J.S.). 
This work was partly supported by MEXT Joint Usage/Research Center on 
Mathematics and Theoretical Physics JPMXP0619217849 (M.Y.) 

\bibliography{Ref-mutopositron}

\begin{thebibliography}{38}
\expandafter\ifx\csname natexlab\endcsname\relax\def\natexlab#1{#1}\fi
\expandafter\ifx\csname bibnamefont\endcsname\relax
  \def\bibnamefont#1{#1}\fi
\expandafter\ifx\csname bibfnamefont\endcsname\relax
  \def\bibfnamefont#1{#1}\fi
\expandafter\ifx\csname citenamefont\endcsname\relax
  \def\citenamefont#1{#1}\fi
\expandafter\ifx\csname url\endcsname\relax
  \def\url#1{\texttt{#1}}\fi
\expandafter\ifx\csname urlprefix\endcsname\relax\def\urlprefix{URL }\fi
\providecommand{\bibinfo}[2]{#2}
\providecommand{\eprint}[2][]{\url{#2}}

\bibitem[{\citenamefont{Marciano and Sanda}(1977)}]{Marciano:1977wx}
\bibinfo{author}{\bibfnamefont{W.~J.} \bibnamefont{Marciano}} \bibnamefont{and}
  \bibinfo{author}{\bibfnamefont{A.~I.} \bibnamefont{Sanda}},
  \bibinfo{journal}{Phys. Lett. B} \textbf{\bibinfo{volume}{67}},
  \bibinfo{pages}{303} (\bibinfo{year}{1977}).

\bibitem[{\citenamefont{Bilenky et~al.}(1977)\citenamefont{Bilenky, Petcov, and
  Pontecorvo}}]{Bilenky:1977du}
\bibinfo{author}{\bibfnamefont{S.~M.} \bibnamefont{Bilenky}},
  \bibinfo{author}{\bibfnamefont{S.~T.} \bibnamefont{Petcov}},
  \bibnamefont{and}
  \bibinfo{author}{\bibfnamefont{B.}~\bibnamefont{Pontecorvo}},
  \bibinfo{journal}{Phys. Lett. B} \textbf{\bibinfo{volume}{67}},
  \bibinfo{pages}{309} (\bibinfo{year}{1977}).

\bibitem[{\citenamefont{Lee and Shrock}(1977)}]{Lee:1977tib}
\bibinfo{author}{\bibfnamefont{B.~W.} \bibnamefont{Lee}} \bibnamefont{and}
  \bibinfo{author}{\bibfnamefont{R.~E.} \bibnamefont{Shrock}},
  \bibinfo{journal}{Phys. Rev. D} \textbf{\bibinfo{volume}{16}},
  \bibinfo{pages}{1444} (\bibinfo{year}{1977}).

\bibitem[{\citenamefont{Lee and MacKenzie}(2021)}]{Lee:2021hnx}
\bibinfo{author}{\bibfnamefont{M.}~\bibnamefont{Lee}} \bibnamefont{and}
  \bibinfo{author}{\bibfnamefont{M.}~\bibnamefont{MacKenzie}}
  (\bibinfo{year}{2021}), \eprint{2110.07093}.

\bibitem[{\citenamefont{Abramishvili et~al.}(2020)\citenamefont{Abramishvili,
  Adamov, Akhmetshin, Allin, Angélique, Anishchik, Aoki, Aznabayev, Bagaturia,
  Ban et~al.}}]{Abramishvili_2020}
\bibinfo{author}{\bibfnamefont{R.}~\bibnamefont{Abramishvili}},
  \bibinfo{author}{\bibfnamefont{G.}~\bibnamefont{Adamov}},
  \bibinfo{author}{\bibfnamefont{R.~R.} \bibnamefont{Akhmetshin}},
  \bibinfo{author}{\bibfnamefont{A.}~\bibnamefont{Allin}},
  \bibinfo{author}{\bibfnamefont{J.~C.} \bibnamefont{Angélique}},
  \bibinfo{author}{\bibfnamefont{V.}~\bibnamefont{Anishchik}},
  \bibinfo{author}{\bibfnamefont{M.}~\bibnamefont{Aoki}},
  \bibinfo{author}{\bibfnamefont{D.}~\bibnamefont{Aznabayev}},
  \bibinfo{author}{\bibfnamefont{I.}~\bibnamefont{Bagaturia}},
  \bibinfo{author}{\bibfnamefont{G.}~\bibnamefont{Ban}}, \bibnamefont{et~al.},
  \bibinfo{journal}{Progress of Theoretical and Experimental Physics}
  \textbf{\bibinfo{volume}{2020}} (\bibinfo{year}{2020}), ISSN
  \bibinfo{issn}{2050-3911},
  \urlprefix\url{http://dx.doi.org/10.1093/ptep/ptz125}.

\bibitem[{\citenamefont{Bartoszek et~al.}(2014)}]{Bartoszek:2014mya}
\bibinfo{author}{\bibfnamefont{L.}~\bibnamefont{Bartoszek}}
  \bibnamefont{et~al.} (\bibinfo{collaboration}{Mu2e}) (\bibinfo{year}{2014}),
  \eprint{1501.05241}.

\bibitem[{\citenamefont{Barlow}(2011)}]{Barlow:2011zza}
\bibinfo{author}{\bibfnamefont{R.~J.} \bibnamefont{Barlow}},
  \bibinfo{journal}{Nucl. Phys. B Proc. Suppl.} \textbf{\bibinfo{volume}{218}},
  \bibinfo{pages}{44} (\bibinfo{year}{2011}).

\bibitem[{\citenamefont{Babu and Mohapatra}(1995)}]{Babu:1995vh}
\bibinfo{author}{\bibfnamefont{K.~S.} \bibnamefont{Babu}} \bibnamefont{and}
  \bibinfo{author}{\bibfnamefont{R.~N.} \bibnamefont{Mohapatra}},
  \bibinfo{journal}{Phys. Rev. Lett.} \textbf{\bibinfo{volume}{75}},
  \bibinfo{pages}{2276} (\bibinfo{year}{1995}), \eprint{hep-ph/9506354}.

\bibitem[{\citenamefont{Weinberg}(1982)}]{Weinberg:1981wj}
\bibinfo{author}{\bibfnamefont{S.}~\bibnamefont{Weinberg}},
  \bibinfo{journal}{Phys. Rev. D} \textbf{\bibinfo{volume}{26}},
  \bibinfo{pages}{287} (\bibinfo{year}{1982}).

\bibitem[{\citenamefont{Sakai and Yanagida}(1982)}]{Sakai:1981pk}
\bibinfo{author}{\bibfnamefont{N.}~\bibnamefont{Sakai}} \bibnamefont{and}
  \bibinfo{author}{\bibfnamefont{T.}~\bibnamefont{Yanagida}},
  \bibinfo{journal}{Nucl. Phys. B} \textbf{\bibinfo{volume}{197}},
  \bibinfo{pages}{533} (\bibinfo{year}{1982}).

\bibitem[{\citenamefont{Hall and Suzuki}(1984)}]{Hall:1983id}
\bibinfo{author}{\bibfnamefont{L.~J.} \bibnamefont{Hall}} \bibnamefont{and}
  \bibinfo{author}{\bibfnamefont{M.}~\bibnamefont{Suzuki}},
  \bibinfo{journal}{Nucl. Phys. B} \textbf{\bibinfo{volume}{231}},
  \bibinfo{pages}{419} (\bibinfo{year}{1984}).

\bibitem[{\citenamefont{Martin}(2010)}]{Martin:1997ns}
\bibinfo{author}{\bibfnamefont{S.~P.} \bibnamefont{Martin}},
  \bibinfo{journal}{Adv. Ser. Direct. High Energy Phys.}
  \textbf{\bibinfo{volume}{21}}, \bibinfo{pages}{1} (\bibinfo{year}{2010}),
  \eprint{hep-ph/9709356}.

\bibitem[{\citenamefont{Barger et~al.}(1989)\citenamefont{Barger, Giudice, and
  Han}}]{Barger:1989rk}
\bibinfo{author}{\bibfnamefont{V.~D.} \bibnamefont{Barger}},
  \bibinfo{author}{\bibfnamefont{G.~F.} \bibnamefont{Giudice}},
  \bibnamefont{and} \bibinfo{author}{\bibfnamefont{T.}~\bibnamefont{Han}},
  \bibinfo{journal}{Phys. Rev.} \textbf{\bibinfo{volume}{D40}},
  \bibinfo{pages}{2987} (\bibinfo{year}{1989}).

\bibitem[{\citenamefont{Amaldi et~al.}(1987)\citenamefont{Amaldi, Bohm, Durkin,
  Langacker, Mann, Marciano, Sirlin, and Williams}}]{Amaldi:1987fu}
\bibinfo{author}{\bibfnamefont{U.}~\bibnamefont{Amaldi}},
  \bibinfo{author}{\bibfnamefont{A.}~\bibnamefont{Bohm}},
  \bibinfo{author}{\bibfnamefont{L.~S.} \bibnamefont{Durkin}},
  \bibinfo{author}{\bibfnamefont{P.}~\bibnamefont{Langacker}},
  \bibinfo{author}{\bibfnamefont{A.~K.} \bibnamefont{Mann}},
  \bibinfo{author}{\bibfnamefont{W.~J.} \bibnamefont{Marciano}},
  \bibinfo{author}{\bibfnamefont{A.}~\bibnamefont{Sirlin}}, \bibnamefont{and}
  \bibinfo{author}{\bibfnamefont{H.~H.} \bibnamefont{Williams}},
  \bibinfo{journal}{Phys. Rev.} \textbf{\bibinfo{volume}{D36}},
  \bibinfo{pages}{1385} (\bibinfo{year}{1987}).

\bibitem[{\citenamefont{Androi? et~al.}(2018)}]{Androic:2018kni}
\bibinfo{author}{\bibfnamefont{D.}~\bibnamefont{Androi?}} \bibnamefont{et~al.}
  (\bibinfo{collaboration}{Qweak}), \bibinfo{journal}{Nature}
  \textbf{\bibinfo{volume}{557}}, \bibinfo{pages}{207} (\bibinfo{year}{2018}),
  \eprint{1905.08283}.

\bibitem[{\citenamefont{Zyla
  et~al.}(2020{\natexlab{a}})}]{ParticleDataGroup:2020ssz}
\bibinfo{author}{\bibfnamefont{P.~A.} \bibnamefont{Zyla}} \bibnamefont{et~al.}
  (\bibinfo{collaboration}{Particle Data Group}), \bibinfo{journal}{PTEP}
  \textbf{\bibinfo{volume}{2020}}, \bibinfo{pages}{083C01}
  (\bibinfo{year}{2020}{\natexlab{a}}).

\bibitem[{\citenamefont{Aaboud et~al.}(2017)}]{ATLAS:2017avc}
\bibinfo{author}{\bibfnamefont{M.}~\bibnamefont{Aaboud}} \bibnamefont{et~al.}
  (\bibinfo{collaboration}{ATLAS}), \bibinfo{journal}{JHEP}
  \textbf{\bibinfo{volume}{11}}, \bibinfo{pages}{195} (\bibinfo{year}{2017}),
  \eprint{1708.09266}.

\bibitem[{\citenamefont{Aaboud et~al.}(2019)}]{ATLAS:2019ebv}
\bibinfo{author}{\bibfnamefont{M.}~\bibnamefont{Aaboud}} \bibnamefont{et~al.}
  (\bibinfo{collaboration}{ATLAS}), \bibinfo{journal}{Eur. Phys. J. C}
  \textbf{\bibinfo{volume}{79}}, \bibinfo{pages}{733} (\bibinfo{year}{2019}),
  \eprint{1902.00377}.

\bibitem[{\citenamefont{Aad et~al.}(2021)}]{ATLAS:2021hza}
\bibinfo{author}{\bibfnamefont{G.}~\bibnamefont{Aad}} \bibnamefont{et~al.}
  (\bibinfo{collaboration}{ATLAS}), \bibinfo{journal}{JHEP}
  \textbf{\bibinfo{volume}{04}}, \bibinfo{pages}{165} (\bibinfo{year}{2021}),
  \eprint{2102.01444}.

\bibitem[{\citenamefont{Vainshtein et~al.}(1975)\citenamefont{Vainshtein,
  Zakharov, and Shifman}}]{Vainshtein:1975sv}
\bibinfo{author}{\bibfnamefont{A.~I.} \bibnamefont{Vainshtein}},
  \bibinfo{author}{\bibfnamefont{V.~I.} \bibnamefont{Zakharov}},
  \bibnamefont{and} \bibinfo{author}{\bibfnamefont{M.~A.}
  \bibnamefont{Shifman}}, \bibinfo{journal}{JETP Lett.}
  \textbf{\bibinfo{volume}{22}}, \bibinfo{pages}{55} (\bibinfo{year}{1975}).

\bibitem[{\citenamefont{Zyla et~al.}(2020{\natexlab{b}})}]{Zyla:2020zbs}
\bibinfo{author}{\bibfnamefont{P.}~\bibnamefont{Zyla}} \bibnamefont{et~al.}
  (\bibinfo{collaboration}{Particle Data Group}), \bibinfo{journal}{PTEP}
  \textbf{\bibinfo{volume}{2020}}, \bibinfo{pages}{083C01}
  (\bibinfo{year}{2020}{\natexlab{b}}).

\bibitem[{\citenamefont{Marciano and Sirlin}(1993)}]{Marciano:1993sh}
\bibinfo{author}{\bibfnamefont{W.~J.} \bibnamefont{Marciano}} \bibnamefont{and}
  \bibinfo{author}{\bibfnamefont{A.}~\bibnamefont{Sirlin}},
  \bibinfo{journal}{Phys. Rev. Lett.} \textbf{\bibinfo{volume}{71}},
  \bibinfo{pages}{3629} (\bibinfo{year}{1993}).

\bibitem[{\citenamefont{Cirigliano and Rosell}(2007)}]{Cirigliano:2007ga}
\bibinfo{author}{\bibfnamefont{V.}~\bibnamefont{Cirigliano}} \bibnamefont{and}
  \bibinfo{author}{\bibfnamefont{I.}~\bibnamefont{Rosell}},
  \bibinfo{journal}{JHEP} \textbf{\bibinfo{volume}{10}}, \bibinfo{pages}{005}
  (\bibinfo{year}{2007}), \eprint{0707.4464}.

\bibitem[{\citenamefont{Bergstrom}(1982)}]{Bergstrom:1982zq}
\bibinfo{author}{\bibfnamefont{L.}~\bibnamefont{Bergstrom}},
  \bibinfo{journal}{Z. Phys. C} \textbf{\bibinfo{volume}{14}},
  \bibinfo{pages}{129} (\bibinfo{year}{1982}).

\bibitem[{\citenamefont{Kitano et~al.}(2002)\citenamefont{Kitano, Koike, and
  Okada}}]{Kitano:2002mt}
\bibinfo{author}{\bibfnamefont{R.}~\bibnamefont{Kitano}},
  \bibinfo{author}{\bibfnamefont{M.}~\bibnamefont{Koike}}, \bibnamefont{and}
  \bibinfo{author}{\bibfnamefont{Y.}~\bibnamefont{Okada}},
  \bibinfo{journal}{Phys. Rev. D} \textbf{\bibinfo{volume}{66}},
  \bibinfo{pages}{096002} (\bibinfo{year}{2002}), \bibinfo{note}{[Erratum:
  Phys.Rev.D 76, 059902 (2007)]}, \eprint{hep-ph/0203110}.

\bibitem[{\citenamefont{Bertl et~al.}(2006)}]{Bertl:2006up}
\bibinfo{author}{\bibfnamefont{W.~H.} \bibnamefont{Bertl}} \bibnamefont{et~al.}
  (\bibinfo{collaboration}{SINDRUM II}), \bibinfo{journal}{Eur. Phys. J.}
  \textbf{\bibinfo{volume}{C47}}, \bibinfo{pages}{337} (\bibinfo{year}{2006}).

\bibitem[{\citenamefont{Suzuki et~al.}(1987)\citenamefont{Suzuki, Measday, and
  Roalsvig}}]{Suzuki:1987jf}
\bibinfo{author}{\bibfnamefont{T.}~\bibnamefont{Suzuki}},
  \bibinfo{author}{\bibfnamefont{D.~F.} \bibnamefont{Measday}},
  \bibnamefont{and} \bibinfo{author}{\bibfnamefont{J.~P.}
  \bibnamefont{Roalsvig}}, \bibinfo{journal}{Phys. Rev. C}
  \textbf{\bibinfo{volume}{35}}, \bibinfo{pages}{2212} (\bibinfo{year}{1987}).

\bibitem[{\citenamefont{Simkovic et~al.}(2001)\citenamefont{Simkovic, Domin,
  Kovalenko, and Faessler}}]{Simkovic:2000ma}
\bibinfo{author}{\bibfnamefont{F.}~\bibnamefont{Simkovic}},
  \bibinfo{author}{\bibfnamefont{P.}~\bibnamefont{Domin}},
  \bibinfo{author}{\bibfnamefont{S.~V.} \bibnamefont{Kovalenko}},
  \bibnamefont{and} \bibinfo{author}{\bibfnamefont{A.}~\bibnamefont{Faessler}},
  \bibinfo{journal}{Part. Nucl. Lett.} \textbf{\bibinfo{volume}{104}},
  \bibinfo{pages}{40} (\bibinfo{year}{2001}), \eprint{hep-ph/0103029}.

\bibitem[{\citenamefont{Vergados}(2002)}]{Vergados:2002pv}
\bibinfo{author}{\bibfnamefont{J.~D.} \bibnamefont{Vergados}},
  \bibinfo{journal}{Phys. Rept.} \textbf{\bibinfo{volume}{361}},
  \bibinfo{pages}{1} (\bibinfo{year}{2002}), \eprint{hep-ph/0209347}.

\bibitem[{\citenamefont{Domin et~al.}(2004)\citenamefont{Domin, Kovalenko,
  Faessler, and Simkovic}}]{Domin:2004tk}
\bibinfo{author}{\bibfnamefont{P.}~\bibnamefont{Domin}},
  \bibinfo{author}{\bibfnamefont{S.}~\bibnamefont{Kovalenko}},
  \bibinfo{author}{\bibfnamefont{A.}~\bibnamefont{Faessler}}, \bibnamefont{and}
  \bibinfo{author}{\bibfnamefont{F.}~\bibnamefont{Simkovic}},
  \bibinfo{journal}{Phys. Rev. C} \textbf{\bibinfo{volume}{70}},
  \bibinfo{pages}{065501} (\bibinfo{year}{2004}), \eprint{nucl-th/0409033}.

\bibitem[{\citenamefont{Divari et~al.}(2002)\citenamefont{Divari, Vergados,
  Kosmas, and Skouras}}]{Divari:2002sq}
\bibinfo{author}{\bibfnamefont{P.~C.} \bibnamefont{Divari}},
  \bibinfo{author}{\bibfnamefont{J.~D.} \bibnamefont{Vergados}},
  \bibinfo{author}{\bibfnamefont{T.~S.} \bibnamefont{Kosmas}},
  \bibnamefont{and} \bibinfo{author}{\bibfnamefont{L.~D.}
  \bibnamefont{Skouras}}, \bibinfo{journal}{Nucl. Phys. A}
  \textbf{\bibinfo{volume}{703}}, \bibinfo{pages}{409} (\bibinfo{year}{2002}),
  \eprint{nucl-th/0203066}.

\bibitem[{\citenamefont{Geib and Merle}(2017)}]{Geib:2016daa}
\bibinfo{author}{\bibfnamefont{T.}~\bibnamefont{Geib}} \bibnamefont{and}
  \bibinfo{author}{\bibfnamefont{A.}~\bibnamefont{Merle}},
  \bibinfo{journal}{Phys. Rev. D} \textbf{\bibinfo{volume}{95}},
  \bibinfo{pages}{055009} (\bibinfo{year}{2017}), \eprint{1612.00452}.

\bibitem[{\citenamefont{Geib et~al.}(2017)\citenamefont{Geib, Merle, and
  Zuber}}]{Geib:2016atx}
\bibinfo{author}{\bibfnamefont{T.}~\bibnamefont{Geib}},
  \bibinfo{author}{\bibfnamefont{A.}~\bibnamefont{Merle}}, \bibnamefont{and}
  \bibinfo{author}{\bibfnamefont{K.}~\bibnamefont{Zuber}},
  \bibinfo{journal}{Phys. Lett. B} \textbf{\bibinfo{volume}{764}},
  \bibinfo{pages}{157} (\bibinfo{year}{2017}), \eprint{1609.09088}.

\bibitem[{\citenamefont{Berryman et~al.}(2017)\citenamefont{Berryman,
  de~Gouv\^ea, Kelly, and Kobach}}]{Berryman:2016slh}
\bibinfo{author}{\bibfnamefont{J.~M.} \bibnamefont{Berryman}},
  \bibinfo{author}{\bibfnamefont{A.}~\bibnamefont{de~Gouv\^ea}},
  \bibinfo{author}{\bibfnamefont{K.~J.} \bibnamefont{Kelly}}, \bibnamefont{and}
  \bibinfo{author}{\bibfnamefont{A.}~\bibnamefont{Kobach}},
  \bibinfo{journal}{Phys. Rev. D} \textbf{\bibinfo{volume}{95}},
  \bibinfo{pages}{115010} (\bibinfo{year}{2017}), \eprint{1611.00032}.

\bibitem[{\citenamefont{Primakoff}(1959)}]{Primakoff:1959fs}
\bibinfo{author}{\bibfnamefont{H.}~\bibnamefont{Primakoff}},
  \bibinfo{journal}{Rev. Mod. Phys.} \textbf{\bibinfo{volume}{31}},
  \bibinfo{pages}{802} (\bibinfo{year}{1959}).

\bibitem[{\citenamefont{Ford and Wills}(1962)}]{FORD1962295}
\bibinfo{author}{\bibfnamefont{K.~W.} \bibnamefont{Ford}} \bibnamefont{and}
  \bibinfo{author}{\bibfnamefont{J.~G.} \bibnamefont{Wills}},
  \bibinfo{journal}{Nuclear Physics} \textbf{\bibinfo{volume}{35}},
  \bibinfo{pages}{295} (\bibinfo{year}{1962}), ISSN \bibinfo{issn}{0029-5582},
  \urlprefix\url{https://www.sciencedirect.com/science/article/pii/002955826290113X}.

\bibitem[{\citenamefont{Kaulard et~al.}(1998)}]{SINDRUMII:1998mwd}
\bibinfo{author}{\bibfnamefont{J.}~\bibnamefont{Kaulard}} \bibnamefont{et~al.}
  (\bibinfo{collaboration}{SINDRUM II}), \bibinfo{journal}{Phys. Lett. B}
  \textbf{\bibinfo{volume}{422}}, \bibinfo{pages}{334} (\bibinfo{year}{1998}).

\bibitem[{\citenamefont{Yeo et~al.}(2017)\citenamefont{Yeo, Kuno, Lee, and
  Zuber}}]{Yeo:2017fej}
\bibinfo{author}{\bibfnamefont{B.}~\bibnamefont{Yeo}},
  \bibinfo{author}{\bibfnamefont{Y.}~\bibnamefont{Kuno}},
  \bibinfo{author}{\bibfnamefont{M.}~\bibnamefont{Lee}}, \bibnamefont{and}
  \bibinfo{author}{\bibfnamefont{K.}~\bibnamefont{Zuber}},
  \bibinfo{journal}{Phys. Rev. D} \textbf{\bibinfo{volume}{96}},
  \bibinfo{pages}{075027} (\bibinfo{year}{2017}), \eprint{1705.07464}.

\end{thebibliography}

\end{document}